\documentclass[11pt,reqno]{extarticle}
\pdfoutput=1
\usepackage{geometry}
\usepackage{graphicx}

\usepackage{anyfontsize}

\usepackage{titlesec}
\usepackage{tocloft}
\usepackage[utf8]{inputenc}
\usepackage[english]{babel}

\usepackage{amssymb}
\usepackage{amsmath}
\usepackage{amsthm}

\usepackage{tikz}
\usetikzlibrary{arrows.meta}

\usepackage{pgfplots}

\usepackage{float}
\usepackage{caption}
\usepackage{subcaption}
\usepackage{scalerel}
\usepackage{enumitem}
\usepackage{hyperref}
\hypersetup{
  colorlinks   = true,
  urlcolor     = blue,
  linkcolor    = blue,
  citecolor   = red
}
\numberwithin{equation}{section}
\allowdisplaybreaks

\newtheorem{theorem}{Theorem}[section]

\newtheorem{definition}[theorem]{Definition}

\newtheorem{remark}[theorem]{Remark}

\newtheorem{proposition}[theorem]{Proposition}

\newtheorem{problem}[theorem]{Problem}

\newcommand{\ed}{\mathrm{d} \makebox[0ex]{}}
\newcommand{\ringg}{\mathring{g} \makebox[0ex]{}}
\newcommand{\dvol}{\mathrm{dvol} \makebox[0ex]{}}

\title{\Large \textsc{Volume-Distance-Ratio Asymptote and Spacetime Inextendibility for FLRW Spacetimes}}
\author{Pengyu Le}
\newcommand{\Address}{{
  \bigskip
  \footnotesize
  \textsc{Beijing Institute of Mathematical Sciences and Applications, Beijing, China}
  
  \textit{E-mail address}: \texttt{pengyu.le@bimsa.cn}
}}
\date{}

\begin{document}

\maketitle

\begin{abstract}
    This paper examines the volume-distance-ratio (VDR) asymptote at the past timelike boundary for Friedmann–Lema\^itre–Robertson–Walker (FLRW) spacetimes. We consider spatially flat FLRW spacetimes with scale factor $a(t) \sim t^{\alpha}$, as well as spatially hyperbolic and spherical FLRW spacetimes with scale factor $a(t) \sim a_0 t^{\alpha}$. Using criteria for spacetime inextendibility based on the VDR asymptote, we investigate the conditions under which these FLRW spacetimes are past inextendible.
\end{abstract}

\tableofcontents

%%%%%%%%%%%%%%%%%%%%%
%%%%%%%%%%%%%%%%%%%%%
%%%%%%%%%%%%%%%%%%%%%
\section{Introduction}

%%%%%%%%%%%%%%%%%%%%%
%%%%%%%%%%%%%%%%%%%%%
\subsection{VDR asymptote under low regularities}
Motivated by the spacetime inextendibility problem in spacetime singularity theory, the author introduced the volume-distance-ratio (VDR) asymptote under low regularities in \cite{Le25}.

\begin{proposition}[{\bf VDR asymptote under $\mathrm{C}^0$ regularity}, see \cite{Le25} Proposition 5.5]
	Let $(M,g)$ be a $(n+1)$-dimensional $\mathrm{C}^0$ globally hyperbolic spacetime and $p\in M$. Let $\{q_k\}\subset I^+(p)$ be a sequence converging to $p$ from a timelike direction. Define the volumes $V_k = \vert I(p,q_k) \vert$ and the distances $d_k = d(p,q_k)$.
	Then the {\bf VDR asymptote} of the chronological diamond $I(p,q_k)$ as $k\rightarrow +\infty$ is
	\begin{align*}
		\frac{V_k}{d_k^{n+1}}
		=
		\frac{2\omega_n}{n+1} \cdot \big( \frac{1}{2} \big)^{n+1} + o(1).
	\end{align*}
\end{proposition}

\begin{proposition}[{\bf VDR asymptote under $\mathrm{C}^{0,1}$ regularity},see \cite{Le25} Proposition 5.6]
	Let $(M,g)$ be a $(n+1)$-dimensional $\mathrm{C}^{0,1}$ globally hyperbolic spacetime and $p\in M$. Let $\gamma$ be a future-directed chronological geodesic emanating from $p$ which satisfies the geodesic equation
	$\ddot{\gamma}^{\kappa} + \Gamma_{\mu\nu}^{\kappa} \dot{\gamma}^{\mu} \dot{\gamma}^{\nu} = 0$,
	where $\Gamma_{\mu\nu}^{\kappa}$ and the differentials $\ed g_{\mu\nu}$ exist along $\gamma$. Define the volume $V(t) = \vert I(p,\gamma(t)) \vert$ and the length $l(t) = L[ \gamma|_{[0,t]} ]$.
	Then the {\bf VDR asymptote} of the chronological diamond $I(p,\gamma(t))$ as $t \rightarrow 0^+$ is
	\begin{align*}
		\frac{V(t)}{l^{n+1}(t)}
		=
		\frac{2\omega_n}{n+1} \cdot \big( \frac{1}{2} \big)^{n+1} + o(1).
	\end{align*}
\end{proposition}

In \cite{Le25}, the author introduced criteria for spacetime inextendibility based on the VDR asymptote and applied to establish inextendibility for the following spacetimes:
\begin{enumerate}[label=\textbullet]
	\item
	      The spatially flat FLRW spacetime $(\mathbb{R}^{n+1}_{t>0},g)$, $n\geq 2$, with asymptotically linear scale factor $a(t) \sim t$, thereby establishing its local $\mathrm{C}^0$ locally null-non-accumulating strongly-causal and $\mathrm{C}^{0,1}$ strongly-causal inextendibility.

	\item
	      The self-similar naked singularity spacetime constructed by Christodoulou in \cite{Chr94}, establishing its local $\mathrm{C}^{0,1}$ strongly-causal inextendibility.
\end{enumerate}

The problem of spacetime inextendibility is closely connected to the strong cosmic censorship conjecture concerning singularities in general relativity \cite{P79} \cite{Chr91}-\cite{Chr09} \cite{D03} \cite{LL18}-\cite{LL22} \cite{LO19a}-\cite{LO19b} \cite{A25} \cite{DL25} \cite{So25} \cite{Gu26} \cite{LS26}. Furthermore, spacetime extensions are intimately tied to the initial value problem and breakdown criteria for rough solutions of Einstein's equations \cite{Cho52} \cite{KR10} \cite{Wq12} \cite{KRS15}. The study of low-regularity spacetime inextendibility also relates to and motivates the investigation of low-regularity Lorentzian geometry \cite{CG12} \cite{FS12} \cite{KSV15} \cite{KSSV15} \cite{Sa16} \cite{GL17} \cite{GGKS18} \cite{Min19} \cite{LLS21} \cite{Gr20} \cite{KOSS22}. Volumes of small causal diamonds in smooth settings was studied primarily in the physics literature \cite{My78} \cite{GS07} \cite{J17} \cite{Wj19}. In \cite{Le23}, the author established a connection between the volume of causal diamonds and a type of isoperimetric inequality. We refer the reader to \cite{Le25} for further discussion on these topics.

In this paper, we further explore the VDR asymptote in spatially flat and hyperbolic FLRW spacetimes and examine their inextendibility. We briefly mention some relevant prior works on spacetime inextendibility:
\begin{enumerate}[label=\textbullet]
	\item
	      \cite{Chr94} ($\mathrm{C}^0$-inextendibility of the naked singularity spacetime within spherically symmetric Lorentzian manifolds).

	\item
	      \cite{Sb18a} \cite{Sb18b} (Schwarzschild spacetime, first demonstration of $\mathrm{C}^0$-inextendibility without symmetry assumptions), \cite{Mie24} ($\mathrm{C}^{0}$-inextendibility for Kasner spacetimes).

	\item
	      \cite{GL17} ($\mathrm{C}^0$-inextendibility of AdS spacetime, $\mathrm{C}^0$-extendibility for a class of spatially hyperbolic $K=-1$ FLRW spacetimes, termed Milne-like, see Remark \ref{rem 1.13}), \cite{Lin20} (Milne-like spatially hyperbolic $K=-1$ FLRW spacetimes).

	\item
	      \cite{GLS18} \cite{GL18} ($\mathrm{C}^0$-inextendibility and $\mathrm{C}^{0,1}$-inextendibility for timelike geodesically complete spacetime), \cite{MS19} (Lorentz-Finsler spaces), \cite{GKS19} (Lorentzian length spaces).

	\item
	      \cite{Sb23} ($\mathrm{C}^0$-inextendibility of a class of spatially spherical and hyperbolic $K=\pm 1$ FLRW spacetimes, see Remarks \ref{rem 1.15}, \ref{rem 1.17}), \cite{Lin26} ($\mathrm{C}^0$-inextendibility of a class of spatially flat $K=0$ FLRW spacetimes).

	\item
	      \cite{Sb22} ($\mathrm{C}_{\text{loc}}^{0,1}$-inextendibility of spherically symmetric weak null singularities constructed in \cite{D03} and cosmological warped product spacetimes, see Remark \ref{rem 1.9}).

	\item
	      \cite{Sb25} ($\mathrm{C}_{\text{loc}}^{0,1}$-inextendibility for weak null singularities constructed in \cite{Lu18} without symmetry assumptions).

\end{enumerate}

%%%%%%%%%%%%%%%%%%%%%
\subsection{Review of VDR-based inextendibility criteria}

Recall the definitions for local spacetime extensions and inextendibility under a given causality condition considered in \cite{Le25}.
\begin{definition}[{\bf Local $\mathrm{C}^{k,\alpha}$ $\mathcal{C}$ (causality condition) extension}, see \cite{Le25} Definition 2.4]
	Let $(M,g)$ be a globally hyperbolic spacetime and $\hat{q}$ be a future/past boundary point.
	A local $\mathrm{C}^{k,\alpha}$ extension of $(M,g)$ at $\hat{q}$ satisfying the following conditions:
	\begin{enumerate}[label=(\alph*)]
		\item
		      A $\mathrm{C}^{k,\alpha}$ spacetime $(\tilde{M}, \tilde{g})$.

		\item
		      $\phi: M \rightarrow \tilde{M}$, a smooth isometric embedding from $M$ to $\tilde{M}$.

		\item
		      $Acc(\{\phi(q_k)\}) \neq \emptyset$: there exists a future/past-ordering chronological sequence $\{q_k\}$ exhausting $\hat{q}$, such that the set of accumulation points of $\{\phi(q_k)\}$ in $\tilde{M}$ is non-empty.
	\end{enumerate}
	If $(\tilde{M},\tilde{g})$ additionally satisfies the causality condition $\mathcal{C}$, then $(\tilde{M}, \tilde{g}, \phi)$ is called a local $\mathrm{C}^{k,\alpha}$ $\mathcal{C}$ extension of $(M,g)$ at $\hat{q}$.
\end{definition}

\begin{definition}[{\bf Local $\mathrm{C}^{k,\alpha}$ ($\mathcal{C}$) inextendibility}, see \cite{Le25} Definition 3.5]
	Let $(M,g)$ be a globally hyperbolic spacetime and $\hat{q}$ be a future/past boundary point.
	$(M,g)$ is locally $\mathrm{C}^{k,\alpha}$ ($\mathcal{C}$) inextendible at $\hat{q}$ if $(M,g)$ admits no local $\mathrm{C}^{k,\alpha}$ ($\mathcal{C}$) extension at $\hat{q}$.
\end{definition}

A key structural property of local null-non-accumulation, introduced in \cite{Le25}, is fundamental for the validity of the VDR asymptote criterion for spacetime inextendibility.
\begin{definition}[{\bf Local null-non-accumulation}, see \cite{Le25} Definition 4.12]
	Let $(M,g)$ be a $\mathrm{C}^2$ globally hyperbolic spacetime and $\hat{q}$ be a future/past boundary point.  Assume that $(\tilde{M}, \tilde{g}, \phi)$ is a local $\mathrm{C}^{k,\alpha}$ extension of $(M,g)$ at $\hat{q}$ along a future/past-ordering chronological sequence $\{q_k\}$ exhausting $\hat{q}$.
	$\phi$ is locally null-non-accumulating at $\hat{q}$ if there exists a future/past-ordering chronological sequence $\{p_k\}\subset I^-(\hat{q})$ exhausting $\hat{q}$, such that $\phi(\hat{q}) \notin Acc(\phi(E_+(q_k,M)))$, where $E^+(q_k,M)$ is the future horismos of $p_k$ in $(M,g)$.
\end{definition}

We now state the VDR asymptote criterion for spacetime inextendibility.
\begin{theorem}[{\bf $\mathrm{C}^0$ locally null-non-accumulating strongly-causal inextendibility}, see \cite{Le25} Theorem 6.1]\label{thm 1.6}
	Let $(M,g)$ be a $\mathrm{C}^2$ globally hyperbolic spacetime and $\hat{q}$ be a future/past boundary point. There exists no local $\mathrm{C}^0$ locally null-non-accumulating strongly-causal extension of $I^-(\hat{q})$ at $\hat{q}$ if for any future/past-ordering chronological sequence $\{ q_k \}$ exhausting $\hat{q}$, we have that
	\begin{align*}
		\frac{V_k}{ d_k^{n+1} } \nrightarrow \frac{2\omega_n}{n+1} \cdot (\frac{1}{2})^{n+1},
	\end{align*}
	where
	$V_k = \vert I(q_k, \hat{q}) \vert$,
	$d_k = d(q_k, \hat{q})$.
\end{theorem}

\begin{theorem}[{\bf $\mathrm{C}^{0,1}$ strongly-causal inextendibility}, see \cite{Le25} Theorem 6.2]\label{thm 1.7}
	Let $(M,g)$ be a $\mathrm{C}^2$ globally hyperbolic spacetime and $\hat{q}$ be a future/past boundary point. There exists no local $\mathrm{C}^{0,1}$ strongly-causal extension of $I^-(\hat{q})$ at $\hat{q}$ if there exists a future/past-directed chronological geodesic curve $\gamma: [t_0,0) \rightarrow M$ exhausting $\hat{q}$, such that as $t\rightarrow 0^-$,
	\begin{align*}
		\frac{V(t)}{ l(t)^{n+1} } \nrightarrow \frac{2\omega_n}{n+1} \cdot (\frac{1}{2})^{n+1},
	\end{align*}
	where
	$V(t) = \vert I(\gamma(t), \hat{q}) \vert$,
	$l(t) = L[\gamma|_{[t,0)}]$.
\end{theorem}

%%%%%%%%%%%%%%%%%%%%
%%%%%%%%%%%%%%%%%%%%
\subsection{VDR asymptote and inextendibility of FLRW spacetimes}

%%%%%%%%%%%%%%%%%%%%
\subsubsection{Spatially flat FLRW spacetimes}

For spatially flat FLRW spacetimes, we derive the following results regarding the VDR asymptote and their inextendibility.
\begin{theorem}
	Consider the spatially flat FLRW spacetime $(M = \mathbb{R}^{n+1}_{t>0},g)$ with scale factor $a(t) \sim t^{\alpha}$.
	\begin{enumerate}[label=\textbullet]
		\item For $n \geq 1$ and $\alpha > 1$, the spacetime admits {\bf no local $\mathrm{C}^0$ locally null-non-accumulating strongly-causal} or {\bf $\mathrm{C}^{0,1}$ strongly-causal extension} at the past timelike boundary point $\mathcal{O}$. The VDR asymptote at $\mathcal{O}$ is {\bf less than} the Minkowski value $\frac{2\omega_n}{n+1} \cdot (\frac{1}{2})^{n+1}$.

		\item For $n = 1$ and $\alpha \in (0,1)$, the spacetime admits {\bf no local $\mathrm{C}^{0,1}$ strongly-causal extension} at any past timelike boundary point. The VDR asymptote along a $\Sigma_t$-orthogonal geodesic towards a past timelike boundary point is {\bf greater than} the Minkowski value $\frac{\omega_1}{4}$.

		\item For $n \geq 2$, there exists a constant $\alpha'_n <1$ such that for $\alpha \in (\alpha'_n,1)$, the spacetime admits {\bf no local $\mathrm{C}^{0,1}$ strongly-causal extension} at any past timelike boundary point. The VDR asymptote along a $\Sigma_t$-orthogonal geodesic towards a past timelike boundary point is {\bf less than} the Minkowski value $\frac{2\omega_n}{n+1} \cdot (\frac{1}{2})^{n+1}$.

		\item For $n \geq 2$ and $\alpha \in (0,1-\log 2]$, the spacetime admits {\bf no local $\mathrm{C}^{0,1}$ strongly-causal extension} at any past timelike boundary point. The VDR asymptote along a $\Sigma_t$-orthogonal geodesic towards a past timelike boundary point is {\bf greater than} the Minkowski value $\frac{2\omega_n}{n+1} \cdot (\frac{1}{2})^{n+1}$.
	\end{enumerate}

	\begin{figure}[h]
		\centering
		\begin{tikzpicture}[scale=7]
			% Draw real number line
			\draw[->] (-0.2,0) -- (1.5,0) node[right] {$\alpha$};

			% Mark points with consistent style
			\foreach \x/\label in {0/{0}, 0.7/{$\alpha'_n$}, 1/{1}}
			\draw (\x,0) -- (\x,0.03) + (0,-0.1) node {\label};

			% Calculate and mark 1 - log2 (natural log)
			\pgfmathsetmacro{\logval}{1 - ln(2)}
			\draw (\logval,0) -- (\logval,0.03) + (0,-0.1) node {$1-\log 2$};

			% Add interval annotations
			% Between 0 and 1-log2
			\node at ({\logval/2}, 0.1) {$> \frac{2\omega_n}{n+1} \cdot (\frac{1}{2} )^{n+1}$};

			% After 1
			\node at (1.3, 0.1) {$< \frac{2\omega_n}{n+1} \cdot (\frac{1}{2} )^{n+1}$};

			% Between 0.9 and 1 (with arrow)
			\node (note) at (0.85, 0.2) {
				\begin{tabular}{l}
					$< \frac{2\omega_n}{n+1} \cdot (\frac{1}{2} )^{n+1},\; n\geq 2$
					\\
					$> \frac{\omega_1}{4},\; n=1$
				\end{tabular}
			};
			\draw[->] (note.south) -- (0.85,0.05);
		\end{tikzpicture}
		\caption{VDR asymptote vs $\alpha$}
		\label{fig 1}
	\end{figure}
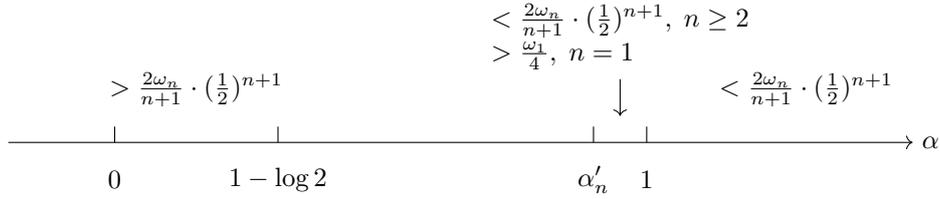
\end{theorem}
\begin{remark}\label{rem 1.9}
	In the case $\alpha \in (0,1)$, the past local $\mathrm{C}^{0,1}$ inextendibility of FLRW spacetimes—regardless of whether the spatial geometry is flat, hyperbolic, or spherical—follows directly from the inextendibility result for cosmological warped product spacetimes in \cite{Sb22}. We present an alternative argument based on the VDR asymptote, which also yields the inextendibility result, albeit under the more restrictive strongly-causal condition, for a certain parameter regime.
\end{remark}

For the exponent $\alpha \in (0,1)$, we establish the existence of a nonzero critical exponent at which the VDR asymptote along a $\Sigma_t$-orthogonal geodesic to a past timelike boundary point is {\bf equal} to the Minkowski value.

\begin{theorem}
	For the spatially flat FLRW spacetime $(M=\mathbb{R}^{n+1}_{t>0}, g_{f})$ with scale factor $a(t) = t^{\alpha}$, $\alpha \in (0,1)$, there exists at least one nonzero critical exponent. Moreover, every nonzero critical exponent lies in $(1-\log 2, 1)$. The spacetime admits no local $\mathrm{C}^{0,1}$ strongly-causal extension at any past timelike boundary point if $\alpha$ is not a critical exponent.
\end{theorem}

Spatially flat FLRW spacetimes corresponding to critical exponents represent interesting geometric objects. We therefore pose the following questions concerning these exponents.

\begin{problem}
\begin{enumerate}[label=\textbullet]
	\item
	      What is the number of nonzero critical exponents? Numerical evidence suggests there is exactly one. Can this be rigorously verified?

	\item
	      How does the critical exponent depend on the spatial dimension $n$? Numerical results indicate that the critical exponent decreases as $n$ increases. Can this behavior be proven?
\end{enumerate}
\end{problem}

%%%%%%%%%%%%%%%%%%%%
\subsubsection{Spatially hyperbolic FLRW spacetimes}

For spatially hyperbolic FLRW spacetimes, we derive the following results regarding the VDR asymptote and their inextendibility.
\begin{theorem}
	For the spatially hyperbolic FLRW spacetime $(M = \mathbb{R}^{n+1}_{t>0}, g)$ with $n\geq 2$ and scale factor $a(t) \sim a_0 t$, the following hold:
	\begin{enumerate}[label=\textbullet]
		\item
		      If $a_0=1$, then the VDR asymptote at the past timelike boundary point is {\bf equal to} the Minkowski value $\frac{2\omega_n}{n+1} \cdot (\frac{1}{2})^{n+1}$.

		\item
		      If $a_0>1$, then the VDR asymptote at the past timelike boundary point is {\bf less than} the Minkowski value $\frac{2\omega_n}{n+1} \cdot (\frac{1}{2})^{n+1}$. Moreover, $(M,g)$ admits {\bf no local $\mathrm{C}^0$ locally null-non-accumulating strongly-causal} or {\bf $\mathrm{C}^{0,1}$ strongly-causal extension} at the past timelike boundary.

		\item
		      If $a_0 \in (0,1)$, then the VDR asymptote at the past timelike boundary point is {\bf greater than} the Minkowski value $\frac{2\omega_n}{n+1} \cdot (\frac{1}{2})^{n+1}$. This asymptote decreases monotonically to $\frac{2\omega_n}{n+1} \cdot (\frac{1}{2})^{n+1}$ as $a_0 \to 1^-$. Furthermore, $(M,g)$ admits {\bf no local $\mathrm{C}^0$ locally null-non-accumulating strongly-causal} or {\bf $\mathrm{C}^{0,1}$ strongly-causal extension} at the past timelike boundary.
	\end{enumerate}
	\begin{remark}\label{rem 1.13}
		In the case $a_0 = 1$, \cite{GL17} defined the Milne-like spacetime and established its past $\mathrm{C}^0$ extendibility.
	\end{remark}

	\begin{figure}[h]
		\centering
		\begin{tikzpicture}[scale=2]
			% Axes
			\draw[->] (0,0) -- (3,0) node[right] {$\alpha$};
			\draw[->] (0,0) -- (0,2) node[above] {$a_0$};

			% Origin label
			\node at (-0.05,-0.05) {$o$};

			% Reference lines
			\draw[dashed] (1,0) node[below] {$1$} -- (1,2) node[above] {$\alpha=1$};
			\draw (0.05,1) -- (0,1) node[left] {$1$};

			% Point (1,1)
			\fill (1,1) circle (1pt) node[above right] {$(1,1)$};

			% Regions
			\node[below] at (2,0.75) {\begin{tabular}{c} \textbf{I} \\ $\alpha > 1$ \end{tabular}};
			\node[below] at (0.5,0.75) {\begin{tabular}{c} \textbf{II} \\ $\alpha \in (0,1)$ \end{tabular}};

			% Boundary indicators
			\draw[<->] (0.2,0.75) -- (0.8,0.75);
			\draw[<->] (1.2,0.75) -- (2.8,0.75);
		\end{tikzpicture}
		\caption{Parameter space $(a_0,\alpha)$}
		\label{fig 2}
	\end{figure}
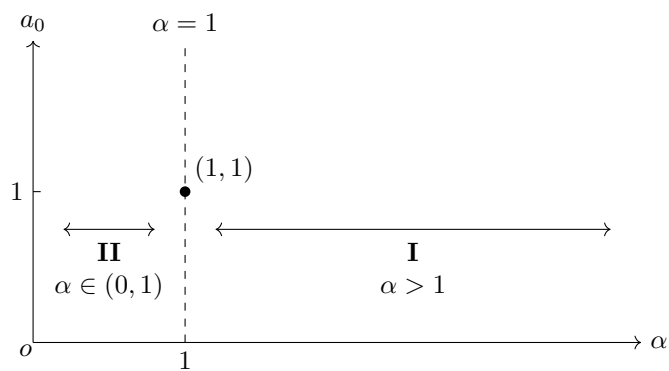
\end{theorem}

\begin{theorem}
	Consider spatially hyperbolic FLRW spacetimes $(M = \mathbb{R}^{n+1}_{t>0},g)$ with $n\geq 2$ and scale factor $a(t) \sim a_0 t^{\alpha}$, $a_0 >0$.
	\begin{enumerate}[label=\textbullet]
		\item For $\alpha \in (0,1)$, the VDR asymptotes along the $\Sigma_t$-orthogonal geodesic coincide with those of the spatially flat FLRW spacetime $(M = \mathbb{R}^{n+1}_{t>0},g_f)$ with scale factor $a(t) = t^{\alpha}$. Consequently, both spacetimes share the same critical exponent set in a given dimension. Moreover, they admit {\bf no local $\mathrm{C}^{0,1}$ strongly-causal extension} at any past timelike boundary point if $\alpha$ is not a critical exponent.

		\item For $\alpha > 1$, the VDR asymptote at the past timelike boundary point is $+\infty$, and $(M,g_h)$ admits {\bf no local $\mathrm{C}^0$ locally null-non-accumulating strongly-causal} or {\bf $\mathrm{C}^{0,1}$ strongly-causal extension} at the past timelike boundary.
	\end{enumerate}
\end{theorem}

\begin{remark}\label{rem 1.15}
	For the case $\alpha\in(0,1)$, we refer to Remark \ref{rem 1.9}. For $n\geq 2,\alpha>1$, and for $n\geq 2,\alpha=1,a_0>1$, the past $\mathrm{C}^0$-inextendibility of the spatially hyperbolic FLRW spacetime can be established using the inextendibility result in \cite{Sb23}. We provide an alternative argument based on the VDR asymptote, which also yields the $\mathrm{C}^0$-inextendibility result at the past timelike boundary for these parameter regimes under the more restrictive conditions of local null-non-accumulation or strong causality.
\end{remark}

%%%%%%%%%%%%%%%%%%%%
\subsubsection{Spatially spherical FLRW spacetimes}

For spatially spherical FLRW spacetimes, we derive the following results regarding the VDR asymptote and their inextendibility.

\begin{theorem}
	Consider spatially spherical FLRW spacetimes $(M = \mathbb{R}_+ \times \mathbb{S}^n,g)$ with $n\geq 1$ and scale factor $a(t) \sim a_0 t^{\alpha}$, $a_0 >0$.
	\begin{enumerate}[label=\textbullet]
		\item For $\alpha > 1$, the VDR asymptote at the past timelike boundary point is $0$, and $(M,g)$ admits {\bf no local $\mathrm{C}^0$ strongly-causal extension} at the past timelike boundary.

		\item For $\alpha = 1$, the VDR asymptote at the past timelike boundary point is {\bf less than} the Minkowski value $\frac{2\omega_n}{n+1}\cdot (\frac{1}{2})^{n+1}$, and $(M,g)$ admits {\bf no local $\mathrm{C}^0$ strongly-causal extension} at the past timelike boundary.

		\item For $n \geq 2$ and $\alpha \in (0,1)$, the VDR asymptotes along the $\Sigma_t$-orthogonal geodesic coincide with those of the spatially flat FLRW spacetime $(M = \mathbb{R}^{n+1}_{t>0},g_f)$ with scale factor $a(t) = t^{\alpha}$. Consequently, both spacetimes share the same critical exponent set in a given dimension. Moreover, they admit {\bf no local $\mathrm{C}^{0,1}$ strongly-causal extension} at any past timelike boundary point if $\alpha$ is not a critical exponent.

		\item For $n = 1$ and $\alpha \in (0,1)$, the VDR asymptotes along the $\Sigma_t$-orthogonal geodesic coincide with those of the spacetime $(M = \mathbb{R}^{2}_{t>0},g_f)$ with scale factor $a(t) = t^{\alpha}$, and are consequently {\bf greater than} the Minkowski value $\frac{\omega_1}{4}$. Moreover, the spacetime admits {\bf no local $\mathrm{C}^{0,1}$ strongly-causal extension} at any past timelike boundary point.
	\end{enumerate}
\end{theorem}

\begin{remark}\label{rem 1.17}
	For the case $\alpha\in(0,1)$, we refer to Remark \ref{rem 1.9}. For $n\geq2,\alpha\geq 1$, the past $\mathrm{C}^0$ inextendibility of the spatially spherical FLRW spacetime can be established using the inextendibility result in \cite{Sb23}. We provide an alternative argument based on the VDR asymptote, which also yields $\mathrm{C}^0$-inextendibility at the past timelike boundary under the more restrictive strongly-causal condition.
\end{remark}

%%%%%%%%%%%%%%%%%%%%
%%%%%%%%%%%%%%%%%%%%
%%%%%%%%%%%%%%%%%%%%
\section{Spatially flat FLRW spacetimes with scale factor $a(t) \sim t^{\alpha}$}

%%%%%%%%%%%%%%%%%%%%
%%%%%%%%%%%%%%%%%%%%
\subsection{Case $\alpha\geq 1$}

In this section, we study a class of spatially flat FLRW spacetimes $(M,g)$ where $M = \mathbb{R}^{n+1}_{t>0}$ and the metric takes the form
\begin{align*}
	 &
	g
	=
	- \ed t^2 + a^2(t) [ (\ed x^1)^2 + \cdots + (\ed x^n)^2]
	=
	- \ed t^2 + a^2(t) ( \ed r^2 + r^2 \ringg ),
	\\
	 &
	a(t) \sim t^{\alpha},
	\quad
	\alpha \geq 1.
\end{align*}
The condition $\int_{t_0}^0  \frac{1}{a(t)} \ed t = \infty$ implies that for any future-directed chronological curve orthogonal to the spatial slices $\Sigma_t$, expressed as
$\gamma(t) = (t, x^1, \cdots , x^n)$,
we obtain
$\cap_{t>0} I^+(\gamma(t)) = M$.

%%%%%%%%%%%%%%%%%%%%
\subsubsection{Conformal compactification and past boundary}

We define
$\tilde{t}
	=
	\int_t^1
	\frac{1}{a(t)} \ed t$,
and the double null coordinates as
\begin{align*}
	 &
	u = \frac{\tilde{t} - r}{2},
	\quad
	v= \frac{\tilde{t} + r}{2}.
\end{align*}
In the double null coordinate system $(u,v, \vartheta)$,
\begin{align*}
	g
	=
	a^2(t) \cdot
	[ - 4 \ed u \ed v
		+  (v-u)^2 \ringg]
	=
	a^2(t) \cdot \eta.
\end{align*}
Considering the transformation
\begin{align*}
	 &
	u = \tan \tilde{u},
	\quad
	v = \tan \tilde{v},
	\quad
	\tilde{u}, \tilde{v} \in (-\frac{\pi}{2} ,\frac{\pi}{2} ),
\end{align*}
we have
\begin{align*}
	g
	=
	\frac{a^2(t)}{\cos^2 \tilde{u} \cos^2 \tilde{v}}
	[ - 4 \ed \tilde{u} \ed \tilde{v}
		+ \sin^2 (\tilde{u} - \tilde{v}) \ringg ].
\end{align*}

From this conformal compactification, the past boundary of $(M,g)$ comprises two components: the past timelike boundary point $\mathcal{O}$ and the past null boundary $\mathcal{PN}$.
\begin{enumerate}[label=\alph*.]
	\item
	      $\mathcal{O}$: $\tilde{u} = \tilde{v} = \frac{\pi}{2}$.

	\item
	      $\mathcal{PN}$: $\tilde{u} \in (-\frac{\pi}{2}, \frac{\pi}{2} )$, $\tilde{v} = \frac{\pi}{2}$, $\vartheta \in \mathbb{S}^{n-1}$.

\end{enumerate}

%%%%%%%%%%%%%%%%%%%%
\subsubsection{Case $\alpha =1$, $n\geq 2$}

We briefly review the calculation of the VDR asymptote for a chronological diamond $I(\mathcal{O},p)$ in the spatially flat FLRW spacetime $(M,g_f)$ with scale factor $a(t) = t$ in \cite{Le25}.
We compare it with the spatially hyperbolic $K=-1$ FLRW spacetime with the metric
\begin{align*}
	g_{h} = - \ed t^2 + t^2 ( \ed r^2 + \sinh^2 r \ringg ),
\end{align*}
which is the Minkowski metric under the transformation
$\tilde{t} = t \cosh r$, $\tilde{r}=t \sinh r$.
Let $p=(t=1,x^1=0,\cdots x^n=0)$. The chronological diamonds $I_f(\mathcal{O},p)$ and $I_h(\mathcal{O},p)$ are identical, simply denoted by $I(\mathcal{O},p)$
\begin{align*}
	I(\mathcal{O},p)
	=
	I^-(p)
	=
	\{ r < \int_t^{1} \frac{1}{t'} \ed t' = - \log t \}.
\end{align*}
The volumes of $I(\mathcal{O},p) \cap \Sigma_t$ and $I(\mathcal{O},p)$
\begin{align*}
	 & \vert I(\mathcal{O},p) \cap \Sigma_t \vert_{g_f}
	=
	t^n \omega_n (- \log t)^n
	<
	t^n \int_0^{- \log t} n \omega_n \sinh^{n-1} r \ed r
	=
	\vert I(\mathcal{O},p) \cap \Sigma_t \vert_{g_h}.
	\\
	 &
	\vert I(\mathcal{O},p) \vert_{g_f}
	=
	\int_0^1 \vert I(\mathcal{O},p) \cap \Sigma_t \vert_{g_f} \ed t
	<
	\int_0^1 \vert I(\mathcal{O},p) \cap \Sigma_t \vert_{g_h} \ed t
	=
	\vert I(\mathcal{O},p) \vert_{g_h},
	\\
	 &
	\vert I(\mathcal{O},p) \vert_{g_h}
	=
	\frac{2\omega_n}{n+1} \cdot (\frac{1}{2})^{n+1}.
\end{align*}
Thus there exist positive constants $\delta_n>0$ such that
\begin{align}
	 &
	\vert I(\mathcal{O},p) \vert_{g_f}
	<
	\vert I(\mathcal{O},p) \vert_{g_h} - \delta_n
	\quad
	\Rightarrow
	\quad
	\frac{\vert I(\mathcal{O},p) \vert_{g_f}}{ d_f(\mathcal{O},p)^{n+1} }
	<
	\frac{2\omega_n}{n+1} \cdot (\frac{1}{2})^{n+1} - \delta_n.
	\label{eqn 2.1}
\end{align}
For the general case, we have the following theorem in \cite{Le25}.

\begin{theorem}[\cite{Le25} Theorem 7.1]\label{thm 2.1}
	The spatially flat FLRW spacetime $(M = \mathbb{R}^{n+1}_{t > 0},g)$ with $n\geq 2$ and
	\begin{align*}
		g
		=
		- \ed t^2 + a^2(t) ( \ed r^2 + r^2 \ringg ),
		\quad
		a(t) \sim t.
	\end{align*}
	admits {\bf no local $\mathrm{C}^0$ locally null-non-accumulating strongly-causal} or {\bf $\mathrm{C}^{0,1}$ strongly-causal extension} at the past timelike boundary point $\mathcal{O}$. The VDR asymptote at $\mathcal{O}$ is {\bf less than} the Minkowski value $\frac{2\omega_n}{n+1} \cdot (\frac{1}{2})^{n+1}$.
\end{theorem}

%%%%%%%%%%%%%%%%%%%%
\subsubsection{Case $\alpha>1$, $n\geq 1$}

We consider the special case $a(t) = t^{\alpha}$ first. The metric takes the form
\begin{align*}
	g_{\alpha,f} =  - \ed t^2 + t^{2\alpha} ( \ed r^2 + r^2 \ringg ).
\end{align*}
By scale-invariance and spatial homogeneity,
it suffices to compute the volume-distance ratio for the chronological diamond $I_{\alpha,f}(\mathcal{O}, p)$, where $p= (t=1, x^1=0, \cdots, x^n=0)$.
We compare these calculations with those for the case $a(t) = t$.
The distance $d_{\alpha,f} (\mathcal{O},p)$ from $p$ to $\mathcal{O}$ is $1$.
The chronological diamond $I_{\alpha,f} (\mathcal{O}, p)$ is given by
\begin{align*}
	I_{\alpha,f} (\mathcal{O}, p)
	=
	I_{\alpha,f}^- (p)
	=
	\{ r < \int_t^1 \frac{1}{t'^{\alpha}} \ed t' = \frac{t^{-\alpha +1} - 1}{\alpha -1} \}.
\end{align*}
Then the volume of $I_{\alpha,f} (\mathcal{O}, p) \cap \Sigma_t$ with respect to the metric $g_{\alpha,f}$ is
\begin{align*}
	\vert I_{\alpha,f} (\mathcal{O}, p) \cap \Sigma_t \vert_{g_{\alpha,f}}
	=
	\omega_n \big( t^{\alpha} \cdot \frac{t^{-\alpha +1} - 1}{\alpha -1} \big)^n
\end{align*}
We have
\begin{align*}
	\vert I_{\alpha,f} (\mathcal{O}, p) \cap \Sigma_t \vert_{g_{\alpha,f}}
	<
	\vert I (\mathcal{O}, p) \cap \Sigma_t \vert_{g_f},
\end{align*}
which is equivalent to
\begin{align*}
	t^{\alpha} \cdot \frac{t^{-\alpha +1} - 1}{\alpha -1}
	<
	t (-\log t )
	\quad
	\Leftrightarrow
	\quad
	1 + \log t^{\alpha-1} < t^{\alpha -1}.
\end{align*}
Therefore,
\begin{align*}
	\vert I_{\alpha,f} (\mathcal{O}, p) \vert_{g_{\alpha,f}}
	<
	\vert I(\mathcal{O}, p) \vert_{g_f}.
\end{align*}
Then there exist constants $\delta_n>0$ such that
\begin{align*}
	\frac{\vert I_{\alpha,f} (\mathcal{O}, p) \vert_{g_{\alpha,f}}}{ [d_{\alpha,f}(\mathcal{O}, p)]^{n+1} }
	<
	\frac{\vert I (\mathcal{O}, p) \vert_{g_f}}{ [d(\mathcal{O}, p)]^{n+1} }
	\left\{
	\begin{aligned}
		 &
		<
		\frac{2\omega_n}{n+1} \cdot (\frac{1}{2})^{n+1} - \delta_n,
		\quad
		n\geq 2,
		\\
		 &
		= \frac{\omega_1}{4},
		\quad
		n=1.
	\end{aligned}
	\right.
\end{align*}
We then extend these calculations to the general case $a(t) \sim |t|^{\alpha}$.

\begin{theorem}\label{thm 2.2}
	The spatially flat FLRW spacetime $(M = \mathbb{R}^{n+1}_{t>0},g)$ with $n\geq 1$ and scale factor
	$a(t) \sim t^{\alpha},
		\alpha > 1$
	admits {\bf no local $\mathrm{C}^0$ locally null-non-accumulating strongly-causal} or {\bf $\mathrm{C}^{0,1}$ strongly-causal extension} at the past timelike boundary point $\mathcal{O}$.	The VDR asymptote at $\mathcal{O}$ is {\bf less than} the Minkowski value $\frac{2\omega_n}{n+1} \cdot (\frac{1}{2})^{n+1}$.
\end{theorem}

\begin{proof}
	By the spatial homogeneity of $g$, we can assume that
	\begin{align*}
		p_k = (t_k, x^1=0, \cdots, x^n=0).
	\end{align*}
	We then have $d(\mathcal{O}, p_k) = t_k$.
	We introduce the radius $r_k(t)$ of $I(\mathcal{O},p_k)$
	\begin{align*}
		I(\mathcal{O},p_k)
		=
		I^-(p_k)
		=
		\{ (t,r): r< r_k(t) \},
		\quad
		r_k (t)
		=
		\int_t^{t_k} \frac{1}{a(t)} \ed t.
	\end{align*}
	Since $a(t) = t^{\alpha} ( 1 + o(1) )$, we have
	\begin{align*}
		r_k(t)
		=
		\int_t^{t_k} \frac{1}{a(t)} \ed t
		=
		(1+ c(t_k, t))
		\cdot
		\frac{ t^{-\alpha + 1} -  t_k^{-\alpha + 1} }{ \alpha - 1}.
	\end{align*}
	Here $c(t',t)$ and $c(t')$ denote arbitrary functions satisfying $c(t',t) = o(1)$ and $c(t') = o(1)$ as $t' \rightarrow 0$.
	The volume of $I(\mathcal{O},p_k) \cap \Sigma_t$ is then
	\begin{align*}
		\vert I(\mathcal{O},p_k) \cap \Sigma_t \vert
		 &
		=
		a^n (t) \omega_n [r_k(t)]^n
		=
		( 1 + c(t_k, t) ) \omega_n \big( t^{\alpha} \frac{ t^{-\alpha + 1} -  t_k^{-\alpha + 1} }{ \alpha - 1} \big)^n.
	\end{align*}
	Setting $t' = t / t_k$, we obtain
	\begin{align*}
		\vert I(\mathcal{O},p_k) \cap \Sigma_t \vert
		=
		( 1 + c(t_k, t) ) t_k^n \cdot \vert I_{\alpha,f}(\mathcal{O}, p) \cap \Sigma_{t'} \vert_{g_{\alpha,f}}.
	\end{align*}
	The volume of $I(\mathcal{O}, p_k)$ is then
	\begin{align*}
		 &
		\vert I(\mathcal{O}, p_k) \vert
		=
		(1+c(t_k)) t_k^{n+1} \cdot \vert I_{\alpha,f} (\mathcal{O}, p) \vert_{g_{\alpha,f}}
		\\
		\Rightarrow
		\quad
		 &
		\frac{\vert I(\mathcal{O}, p_k) \vert}{ [d(\mathcal{O}, p_k)]^{n+1}}
		=
		( 1 + c(t_k) )
		\frac{\vert I_{\alpha,f} (\mathcal{O}, p) \vert_{g_{\alpha,f}}}{ [d_{\alpha,f}(\mathcal{O}, p)]^{n+1} }.
	\end{align*}
	Therefore the VDR asymptote at $\mathcal{O}$ is
	\begin{align*}
		\lim_{t_k \rightarrow0}
		\frac{\vert I(\mathcal{O}, p_k) \vert}{ [d(\mathcal{O}, p_k)]^{n+1}}
		\leq
		\frac{\vert I_{\alpha,f} (\mathcal{O}, p) \vert_{g_{\alpha,f}}}{ [d_{\alpha,f}(\mathcal{O}, p)]^{n+1} }
		<
		\frac{2\omega_n}{n+1} \cdot (\frac{1}{2})^{n+1}.
	\end{align*}
	Hence, by Theorems \ref{thm 1.6} and \ref{thm 1.7}, the result follows.
\end{proof}

%%%%%%%%%%%%%%%%%%%%
%%%%%%%%%%%%%%%%%%%%
\subsection{Case $\alpha \in (0,1)$}

In this section, we examine another class of spatially flat FLRW spacetimes $(M,g)$ characterized by a scale factor $a(t) \sim t^{\alpha}$ with $0< \alpha <1$.

Unlike the case $\alpha \geq 1$, the integral $\int_0^1 \frac{1}{a(t)} \ed t$ is finite. Consequently, for the future-directed timelike curve $\gamma_0(t) = (t, x^1=0, \cdots, x^n = 0)$ orthogonal to the spatial slices $\Sigma_t$, we find $\cap_{t >0} I^+(t) = \{ (t,x): |x| < \int_0^t \frac{1}{a(t)} \ed t \}$.

%%%%%%%%%%%%%%%%%%%%
\subsubsection{Past timelike boundary}As in the case $\alpha \geq 1$, we define
$
	\tilde{t}
	=
	\int_0^t
	\frac{1}{a(t)} \ed t < \infty$.
Then $\ed \tilde{t} = \frac{1}{a(t)} \ed t$, and in the coordinates $(\tilde{t}, r, \vartheta \in \mathbb{S}^{n-1})$, the metric becomes
\begin{align*}
	g
	=
	a(t)^2 ( - \ed \tilde{t}^2 + \ed r^2 + r^2 \ringg)
	=
	a(t)^2 \eta.
\end{align*}
Through this coordinate transformation and the compactification of Minkowski spacetime, we observe that $(M,g)$ possesses a past timelike boundary $\mathcal{PT}$ and an empty past null boundary, given by
\begin{align*}
	\mathcal{PT}
	=
	\{ \tilde{t} = 0 \}
	=
	\{ t = 0 \}.
\end{align*}

%%%%%%%%%%%%%%%%%%%%
\subsubsection{Case $\alpha \in (0,1)$, $n=1$}
\label{sec 2.2.2}

We examine the specific metric $g_{\alpha,f}$ with scale factor $a(t) = t^{\alpha}, \alpha \in [0,1]$.
Note that for $\alpha=0$ and $\alpha=1$, the metric $g_{\alpha,f}$ corresponds to the 2-dimensional Minkowski metric. To compute the volume of the chronological diamond $I_{\alpha,f}(o,p)$ with $p=(t=1, x=0)$, we define the radius functions
\begin{align*}
	r_{\alpha,f}^+(t)
	=
	\int_0^t t^{-\alpha} \ed t
	=
	\frac{t^{1-\alpha}}{1-\alpha},
	\quad
	r_{\alpha,f}^-(t)
	=
	\int_t^1 t^{-\alpha} \ed t
	=
	\frac{1- t^{1-\alpha}}{1-\alpha}.
\end{align*}
The chronological diamond is then given by
\begin{align}
	 &
	I_{\alpha,f}(o,p)
	=
	\{ (t,x): |x| < r_{\alpha,f}(t) \},
	\nonumber
	\\
	 &
	r_{\alpha,f}(t)
	=
	\min \{ r_{\alpha,f}^+(t), r_{\alpha,f}^-(t) \}
	=
	\left\{
	\begin{aligned}
		 &
		\frac{t^{1-\alpha}}{1-\alpha},
		 &   &
		t \in (0, (\frac{1}{2})^{\frac{1}{1-\alpha}} ]
		\\
		 &
		\frac{1-t^{1-\alpha}}{1-\alpha},
		 &   &
		t \in [ (\frac{1}{2})^{\frac{1}{1-\alpha}}, 1 ]
	\end{aligned}
	\right.
	\label{eqn 2.2}
\end{align}
The volume of $I_{\alpha,f}(o,p)$ becomes
\begin{align*}
	\vert I_{\alpha,f}(o,p) \vert
	 &
	=
	\int_0^1 \vert I_{\alpha,f} (o,p) \cap \Sigma_{t} \vert \ed t
	=
	\int_0^1 \omega_1 t^{\alpha} r_{\alpha, h}(t)  \ed t
	\\
	 &
	=
	\int_0^{(\frac{1}{2})^{\frac{1}{1-\alpha}}}
	\frac{\omega_1}{1-\alpha} t \ed t
	+
	\int_{(\frac{1}{2})^{\frac{1}{1-\alpha}}}^1
	\frac{\omega_1}{1-\alpha} ( t^{\alpha}- t ) \ed t
	=
	\frac{\omega_1}{(1+\alpha)} \big[ \frac{1}{2} - (\frac{1}{2} )^{\frac{2}{1-\alpha}}  \big].
\end{align*}
To show that $\frac{1}{(1+\alpha)} \big[ \frac{1}{2} - (\frac{1}{2} )^{\frac{2}{1-\alpha}}  \big] > \frac{1}{4}$ for $\alpha \in (0,1)$, we note the equivalent inequality $\frac{4}{1-\alpha} < 2^{\frac{2}{1-\alpha}}$ follows from the fact that $2^x > 2x$ for $x\in (2,\infty)$. Hence, we have established that for $\alpha \in (0,1)$,
\begin{align*}
	\frac{\vert I_{\alpha,f}(o,p) \vert}{ [d_{\alpha,f}(o,p)]^2 }
	>
	\frac{\omega_1}{4}.
\end{align*}
We now extend these calculations to the general case.

\begin{theorem}\label{thm 2.3}
	The spatially spacetime $(M = \mathbb{R}^{2}_{t>0},g)$ with scale factor
	$a(t) \sim t^{\alpha},
		\alpha \in (0,1)$,
	admits {\bf no $\mathrm{C}^{0,1}$ strongly-causal extension} at any past timelike boundary point. Moreover, the VDR asymptote along a $\Sigma_t$-orthogonal geodesic approaching a past timelike boundary point is {\bf greater than} the Minkowski value $\frac{\omega_1}{4}$.
\end{theorem}

\begin{proof}
	We estimate the VDR asymptote along the geodesic $\gamma(t) = (t, x=0)$. We show that
	\begin{align*}
		\lim_{t \rightarrow 0^-} \frac{ V(t) }{ l^{2}(t) }
		>
		\frac{\omega_1}{4}.
	\end{align*}
	The proof follows by straightforward modification of the arguments used in Theorem \ref{thm 2.2}, so we omit the details.
\end{proof}

%%%%%%%%%%%%%%%%%%%%
\subsubsection{Case $\alpha$ near $1$, $n\geq 2$}

By a straightforward continuity argument, for $n \geq 2$ and $\alpha$ sufficiently close to $1$, the VDR asymptote along the timelike geodesic $\gamma(t) = (t, x^1=0, \cdots, x^n=0)$ approaches that of the linear scale factor $a(t)=t$, which is smaller than the corresponding Minkowski value. Therefore, Theorem \ref{thm 1.7} implies the associated $\mathrm{C}^{0,1}$-inextendibility.

\begin{theorem}\label{thm 2.4}
	There exists a constant $\alpha'_n <1$, such that the spatially flat FLRW spacetime $(M = \mathbb{R}^{n+1}_{t>0},g)$ with $n\geq 2$ and scale factor
	$a(t) \sim t^{\alpha},
		\alpha'_n < \alpha <1$,
	admits {\bf no local $\mathrm{C}^{0,1}$ strongly-causal extension} at any past timelike boundary point. The VDR asymptote along the $\Sigma_t$-orthogonal geodesic towards a past timelike boundary point is {\bf less than} the Minkowski value $\frac{2\omega_n}{n+1} \cdot (\frac{1}{2})^{n+1}$.
\end{theorem}

\begin{proof}
	We compute the VDR asymptote along the geodesic $\gamma(t) = (t, x^1=0, \cdots, x^n=0)$, which converges to a past boundary point $o$ as $t \rightarrow 0^+$; here the distance $l(t)$ from $\gamma(t)$ to $o$ is simply $l(t) = t$. To estimate the volume of the chronological diamond $I(\gamma(t_0), o)$, we define the radius functions
	\begin{align*}
		 &
		I^+(o)
		=
		\{ (t, r): r< r^+(t) \},
		\quad
		r^+(t)
		=
		\int_0^t \frac{1}{a(t)} \ed t,
		\\
		 &
		I^-(\gamma(t_0))
		=
		\{ (t, r): r< r_{t_0}^-(t) \},
		\quad
		r_{t_0}^-(t)
		=
		\int_t^{t_0} \frac{1}{a(t)} \ed t,
		\\
		 &
		I(\gamma(t_0), o)
		=
		\{ (t, r): r< \min \{ r^+(t), r_{t_0}^-(t) \} \}.
	\end{align*}
	Since $V(t_0)=|I(o,\gamma(t_0))| < |I^-(\gamma(t_0))|$, we obtain
	\begin{align*}
		\limsup_{t_0 \rightarrow 0} \frac{V(t_0)}{[l(t_0)]^{n+1}}
		\leq
		\lim_{t_0 \rightarrow 0} \frac{|I^-(\gamma(t_0))|}{[l(t_0)]^{n+1}}
		\leq
		\int_0^1 \omega_n  {t}^{n\alpha}  \big( \frac{1 - {t}^{1-\alpha}}{1-\alpha} \big)^n \ed t,
	\end{align*}
	where the last inequality follows from a straightforward adaptation of the argument in Theorem \ref{thm 2.2}. Note that for $\alpha<1$, we have $\frac{1 - {t}^{1-\alpha}}{1-\alpha} < - \log t$ with $\lim_{\alpha \rightarrow 1^-} \frac{1 - {t}^{1-\alpha}}{1-\alpha} = - \log t$, so by Lebesgue's dominated convergence theorem,
	\begin{align*}
		\lim_{\alpha \rightarrow 1^-} \int_0^1 \omega_n  {t}^{n\alpha}  \big( \frac{1 - {t}^{1-\alpha}}{1-\alpha} \big)^n \ed t
		=
		\int_0^1 \omega_n  {t}^n \big(- \log t \big)^n \ed t
		<
		\frac{2\omega_n}{n+1} \cdot (\frac{1}{2})^{n+1} - \delta_n,
	\end{align*}
	with the final inequality following from \eqref{eqn 2.1}. Therefore, there exists $\alpha'_n<1$ sufficiently close to $1$ such that for all $\alpha \in (\alpha'_n,1)$,
	\begin{align*}
		\limsup_{t_0 \rightarrow 0} \frac{V(t_0)}{l^{n+1}(t_0)}
		<
		\frac{2\omega_n}{n+1} \cdot (\frac{1}{2})^{n+1}  - \frac{1}{2}\delta_n.
	\end{align*}
	Hence, by Theorem \ref{thm 1.7}, the result follows.
\end{proof}

%%%%%%%%%%%%%%%%%%%%
\subsubsection{Case $0< \alpha \ll 1$, $n \geq 2$}

In this section, we examine positive values of $\alpha$ near $0$. Note that when $\alpha=0$ (corresponding to $a(t) = 1$), the spatially flat FLRW metric reduces to the Minkowski metric. We compute the VDR asymptote along the curve $\gamma(t) = ( t, |x|=0)$ for the specific metric $g_{\alpha,f}$ with scale factor $a(t) = t^{\alpha}$.
By scale invariance and spatial homogeneity, it suffices to compute the volume of the chronological diamond $I_{\alpha,f}(o,p)$ with $p=(t=1, |x|=0)$. From \eqref{eqn 2.2}, the volume of $I_{\alpha,f}(o,p)$ is given by
\begin{align*}
	\vert I_{\alpha,f}(o,p) \vert
	=
	\int_0^{(\frac{1}{2})^{\frac{1}{1-\alpha}}}
	\omega_n \big( \frac{1}{1-\alpha} \big)^n t^{n} \ed t
	+
	\int_{(\frac{1}{2})^{\frac{1}{1-\alpha}}}^1
	\omega_n \big( \frac{1}{1-\alpha} \big)^n ( t^{\alpha}- t)^n \ed t.
\end{align*}
Differentiating $\vert I_{\alpha,f}(o,p) \vert$ yields
\begin{align*}
	\frac{\ed}{\ed \alpha} \vert I_{\alpha,f}(o,p) \vert
	 &
	=
	\int_0^{(\frac{1}{2})^{\frac{1}{1-\alpha}}}
	n \omega_n \big( \frac{1}{1-\alpha} \big)^{n+1} {t}^{n} \ed t
	\\
	 & \phantom{=}
	+
	\int_{(\frac{1}{2})^{\frac{1}{1-\alpha}}}^1
	n \omega_n \big( \frac{1}{1-\alpha} \big)^{n+1} ( {t}^{\alpha} - t )^n \ed t
	\\
	 & \phantom{=}
	+
	\int_{(\frac{1}{2})^{\frac{1}{1-\alpha}}}^1
	n \omega_n \big( \frac{1}{1-\alpha} \big)^n {t}^{\alpha} ( {t}^{\alpha}- t )^{n-1} \log t \ed t.
\end{align*}
Evaluating at $\alpha=0$ gives
\begin{align*}
	\frac{\ed}{\ed \alpha} \vert I_{\alpha,f}(o,p) \vert \Big|_{\alpha=0}
	 &
	=
	\int_0^{\frac{1}{2}}
	n \omega_n {t}^{n} \ed t
	+
	\int_{\frac{1}{2}}^1
	n \omega_n ( 1 - t )^n \ed t
	%\\
	%	&\phantom{=}
	+
	\int_{\frac{1}{2}}^1
	n \omega_n ( 1 - t )^{n-1} \log t \ed t
	\\
	 &
	=
	n \omega_n
	\int_{\frac{1}{2}}^1
	( 1 - t )^{n-1} \big( 2  - 2 t +  \log t \big) \ed t > 0,
\end{align*}
which holds because $2 - 2 t + \log t > 0$ for $ t \in (\frac{1}{2}, 1)$. Hence, for sufficiently small $\alpha>0$,
\begin{align*}
	\vert I_{\alpha,f}(o,p) \vert
	>
	\vert I_{0,f}(o,p) \vert
	=
	\frac{2\omega_n}{n+1} \cdot (\frac{1}{2})^{n+1}.
\end{align*}
We extend the calculations to the general case.

\begin{theorem}
	The spatially flat FLRW spacetime $(M = \mathbb{R}^{n+1}_{t>0},g)$ with $n\geq 2$ and scale factor
	$a(t) \sim t^{\alpha},
		\alpha'_n < \alpha <1$,
	admits {\bf no local $\mathrm{C}^{0,1}$ strongly-causal extension} at any past timelike boundary point for sufficiently small positive $\alpha$, depending on $n$. The VDR asymptote along a $\Sigma_t$-orthogonal geodesic towards a past timelike boundary point is {\bf greater than} the Minkowski value $\frac{2\omega_n}{n+1} \cdot (\frac{1}{2})^{n+1}$.
\end{theorem}

\begin{proof}
	We estimate the VDR asymptote along the geodesic $\gamma(t) = (t, |x|=0)$. We show that
	\begin{align*}
		\lim_{t \rightarrow 0^+} \frac{ V(t) }{ l^{n+1}(t) }
		>
		\frac{2\omega_n}{n+1} \cdot (\frac{1}{2})^{n+1}.
	\end{align*}
	The proof follows by straightforward modification of the arguments used in Theorem \ref{thm 2.2}, so we omit the details.
\end{proof}

%%%%%%%%%%%%%%%%%%%%
\subsubsection{Case $\alpha \in (0,1-\log 2]$, $n\geq 2$}

Following the preceding calculation, we examine the derivative $\frac{\ed}{\ed \alpha} \vert I_{\alpha,f}(o,p) \vert$ in greater detail. We introduce the following change of variables
\begin{align*}
	\big( 0, (\frac{1}{2})^{\frac{1}{1-\alpha}} \big)
	\leftrightarrow
	\big( (\frac{1}{2})^{\frac{1}{1-\alpha}}, 1 \big),
	\quad
	t \leftrightarrow t^{\alpha} - t.
\end{align*}
We then obtain
$\ed t
	\leftrightarrow
	( \alpha t^{\alpha-1} - 1 ) \ed t$,
yielding
\begin{align*}
	n \omega_n \big( \frac{1}{1-\alpha} \big)^{n+1} {t}^{n} \ed t
	\leftrightarrow
	- n \omega_n \big( \frac{1}{1-\alpha} \big)^{n+1} (t^{\alpha} - t)^{n} ( - \alpha t^{\alpha-1} + 1 ) \ed t.
\end{align*}
Therefore
\begin{align*}
	\frac{\ed}{\ed \alpha} \vert I_{\alpha,f}(o,p) \vert
	 &
	=
	\int_{(\frac{1}{2})^{\frac{1}{1-\alpha}}}^1
	n \omega_n \big( \frac{1}{1-\alpha} \big)^{n+1} ( {t}^{\alpha}- t )^{n-1}
	\\
	 & \phantom{= \int_{(\frac{1}{2})^{\frac{1}{1-\alpha}}}^1 n \omega_n}
	\cdot
	\big[
		(2 - \alpha {t}^{\alpha-1}) ( {t}^{\alpha} - t )
		+
		(1-\alpha) {t}^{\alpha} \log t
		\big]
	\ed t.
\end{align*}

\begin{theorem}
	The spatially flat FLRW spacetime $(M = \mathbb{R}^{n+1}_{t>0},g)$ with $n\geq 2$ and scale factor $a(t) \sim |t|^{\alpha}$,
	admits {\bf no local $\mathrm{C}^{0,1}$ strongly-causal extension} at any past timelike boundary point for $0<\alpha \leq \alpha_0$ where
	\begin{align*}
		\alpha_0 = 1- \log 2 \approx 0.307.
	\end{align*}
	The VDR asymptote along a $\Sigma_t$-orthogonal geodesic towards a past timelike boundary point is {\bf greater than} the Minkowski value $\frac{2\omega_n}{n+1} \cdot (\frac{1}{2})^{n+1}$.
\end{theorem}

\begin{proof}
	The preceding calculation shows that
	\begin{align*}
		\frac{\ed}{\ed \alpha} \vert I_{\alpha,f}(o,p) \vert > 0
	\end{align*}
	if
	$(2 - \alpha t^{\alpha-1}) ( t^{\alpha} - t )
		+
		(1-\alpha) t^{\alpha} \log t
		>
		0$,
	for $t \in \big( (\frac{1}{2})^{\frac{1}{1-\alpha}}, 1 \big)$,
	which is equivalent to
	\begin{align*}
		\alpha
		<
		\frac{2}{t^{\alpha-1}} - \frac{\log t^{\alpha-1}}{t^{\alpha-1} -1 },
	\end{align*}
	noting that $t^{\alpha-1} \in (1, 2)$.
	We define the function
	\begin{align*}
		f(u) = \frac{2}{u} - \frac{\log u}{u-1}.
	\end{align*}
	Direct calculation shows that
	\begin{align*}
		f'(u)
		 &
		=
		- \frac{2}{u^2}
		- \frac{1}{u(u-1)}
		+ \frac{\log u}{(u-1)^2}
		\\
		 &
		<
		- \frac{2}{u^2}
		- \frac{1}{u(u-1)}
		+ \frac{1}{u-1}
		=
		\frac{u-2}{u^2} < 0,
		\quad
		u\in (1,2)
	\end{align*}
	so $f'(u) <0 $ for $u = t^{\alpha-1} \in (1, 2)$, and thus for $\alpha \in [0,\alpha_0]$ we have
	\begin{align*}
		\alpha
		\leq
		\alpha_0
		=
		f(2)
		<
		f(u),
		\quad
		u \in (1, 2).
	\end{align*}
	Therefore,
	\begin{align*}
		\frac{\ed}{\ed \alpha} \vert I_{\alpha,f}(o,p) \vert
		>
		0
		\quad
		\Rightarrow
		\quad
		\vert I_{\alpha,f}(o,p) \vert
		>
		\vert I_{0,f}(o,p) \vert.
	\end{align*}
	The general case follows from the same argument as in the proofs of Theorem \ref{thm 2.2}, so we omit the details.
\end{proof}

%%%%%%%%%%%%%%%%%%%%
\subsubsection{Existence of critical exponent for $n\geq 2$}

For the spatially flat FLRW spacetime $(M=\mathbb{R}^{n+1}_{t>0}, g_{\alpha,f})$ with $n\geq 2$ and scale factor $a(t) = t^{\alpha}$, the volume-diameter ratio (VDR) is defined as a function $R$ of $\alpha$
\begin{align*}
	R_n (\alpha)
	=
	\frac{\vert I_{\alpha,f}(o,p) \vert_{g_{\alpha,f}}}{ [d(o,p)]^{n+1} }
	=
	\vert I_{\alpha,f}(o,p) \vert_{g_{\alpha,f}},
\end{align*}
where $p=(t=1, x^1=0, \cdots , x^n=0)$.
We then have
\begin{align*}
	 &
	\alpha = 0,
	\quad
	R_n (0) = \frac{2\omega_n}{n+1} \cdot (\frac{1}{2})^{n+1},
	\\
	 &
	\alpha \in (0, a_0],
	\quad
	R_n (\alpha) > \frac{2\omega_n}{n+1} \cdot (\frac{1}{2})^{n+1},
	\\
	 &
	\alpha \in ( \alpha'_n, 1),
	\quad
	R_n (\alpha) < \frac{2\omega_n}{n+1} \cdot (\frac{1}{2})^{n+1}.
\end{align*}

\begin{definition}\label{def 2.7}
	For the spatially flat FLRW spacetime $(M=\mathbb{R}^{n+1}_{t>0}, g_{\alpha,f})$ with scale factor $a(t)=t^{\alpha}$, we call an exponent $\alpha$ critical if
	\begin{align*}
		R_n (\alpha) = \frac{2\omega_n}{n+1} \cdot (\frac{1}{2})^{n+1}.
	\end{align*}
\end{definition}

\begin{theorem}\label{thm 2.8}
	The spatially flat FLRW spacetime $(M = \mathbb{R}^{n+1}_{t>0},g)$, with $n\geq 2$ and scale factor $a(t) \sim t^{\alpha}$,
	admits {\bf no local $\mathrm{C}^{0,1}$ strongly-causal extension} at any past timelike boundary point if $\alpha$ is not a critical exponent.
\end{theorem}
\begin{proof}
	It follows from the definition of the critical exponent and Theorem \ref{thm 1.7}.
\end{proof}

\begin{theorem}\label{thm 2.9}
	For the spatially flat FLRW spacetime $(M=\mathbb{R}^{n+1}_{t>0}, g_{\alpha,f})$ with $n\geq 2$ and scale factor $a(t) = t^{\alpha}$, $\alpha \in [0,1)$, there exists at least one nonzero critical exponent. Moreover, the nonzero critical exponent must lie in $(1-\log 2, 1)$.
\end{theorem}
\begin{proof}
	It follows from the continuity of the function $R_n (\alpha)$.
\end{proof}

\begin{remark}\label{rem 2.10}
	Numerical calculations indicate the existence of a unique nonzero critical exponent $\alpha_n$ (see Figure \ref{fig 3}). Furthermore, these calculations suggest that the critical exponent $\alpha_n$ decreases as $n$ increases (see Figure \ref{fig 4}). It would be interesting to investigate whether the geometry of the spatially flat FLRW spacetime corresponding to this critical exponent possesses special properties.
	\begin{figure}[h]
		\centering
		\includegraphics[width=0.45\textwidth]{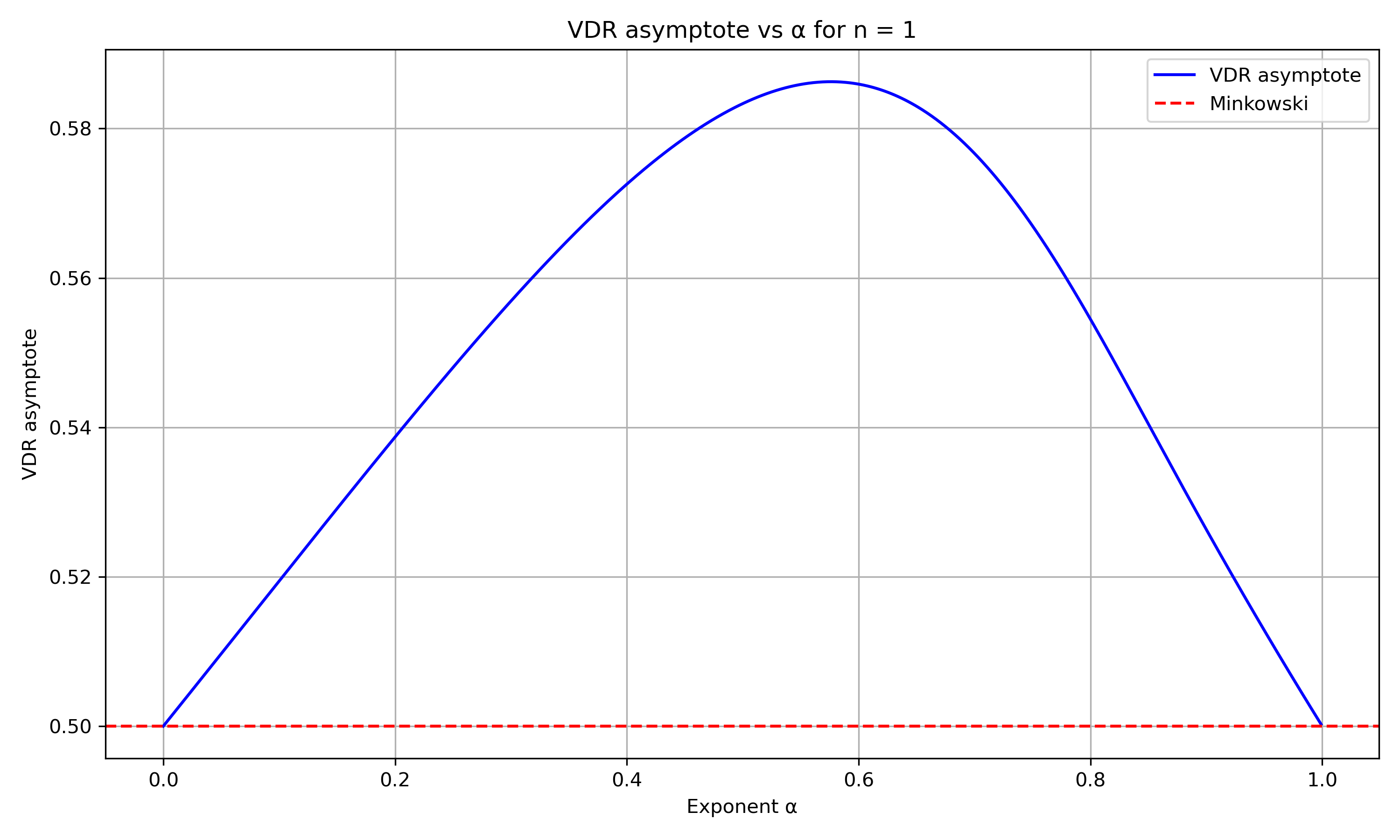}
		\includegraphics[width=0.45\textwidth]{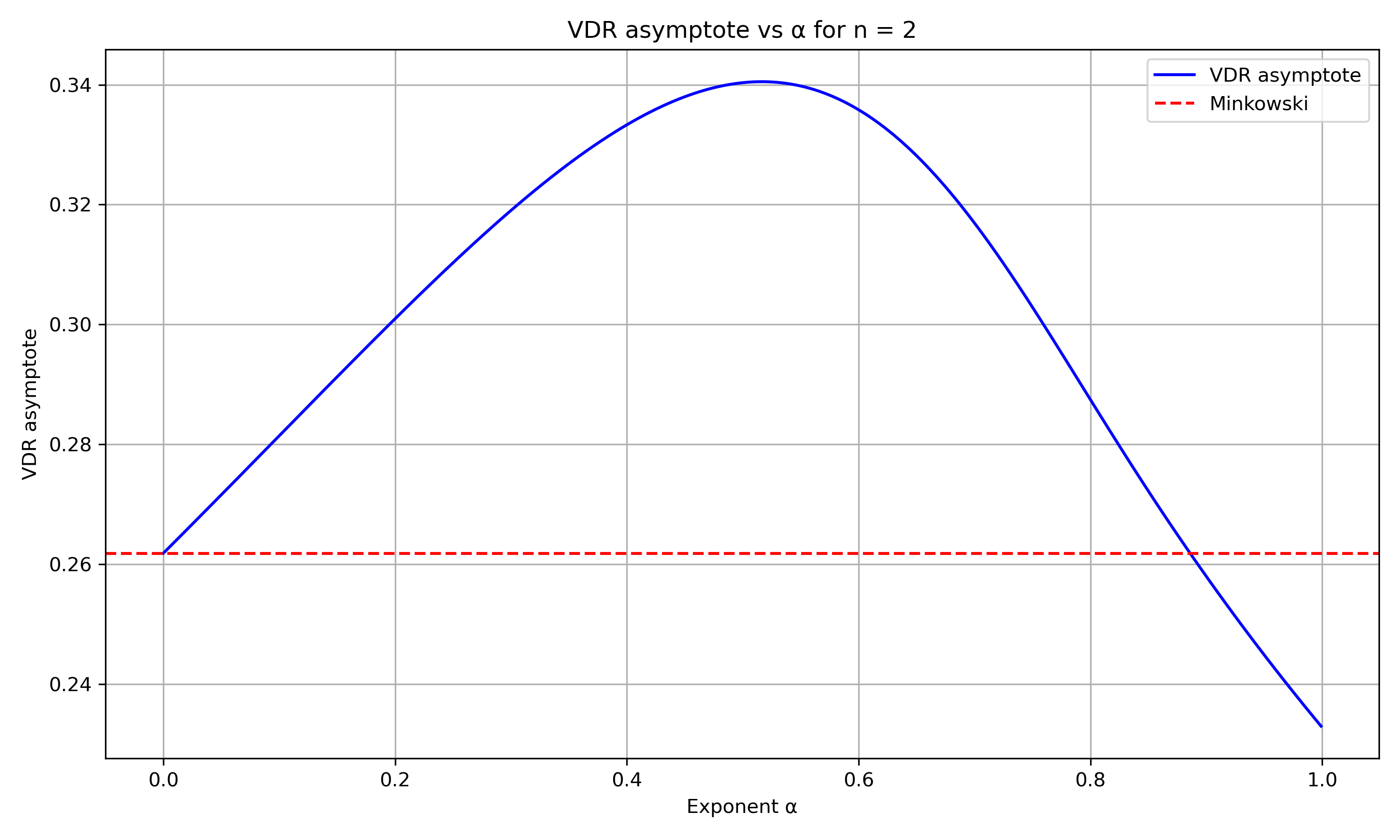}
		\includegraphics[width=0.45\textwidth]{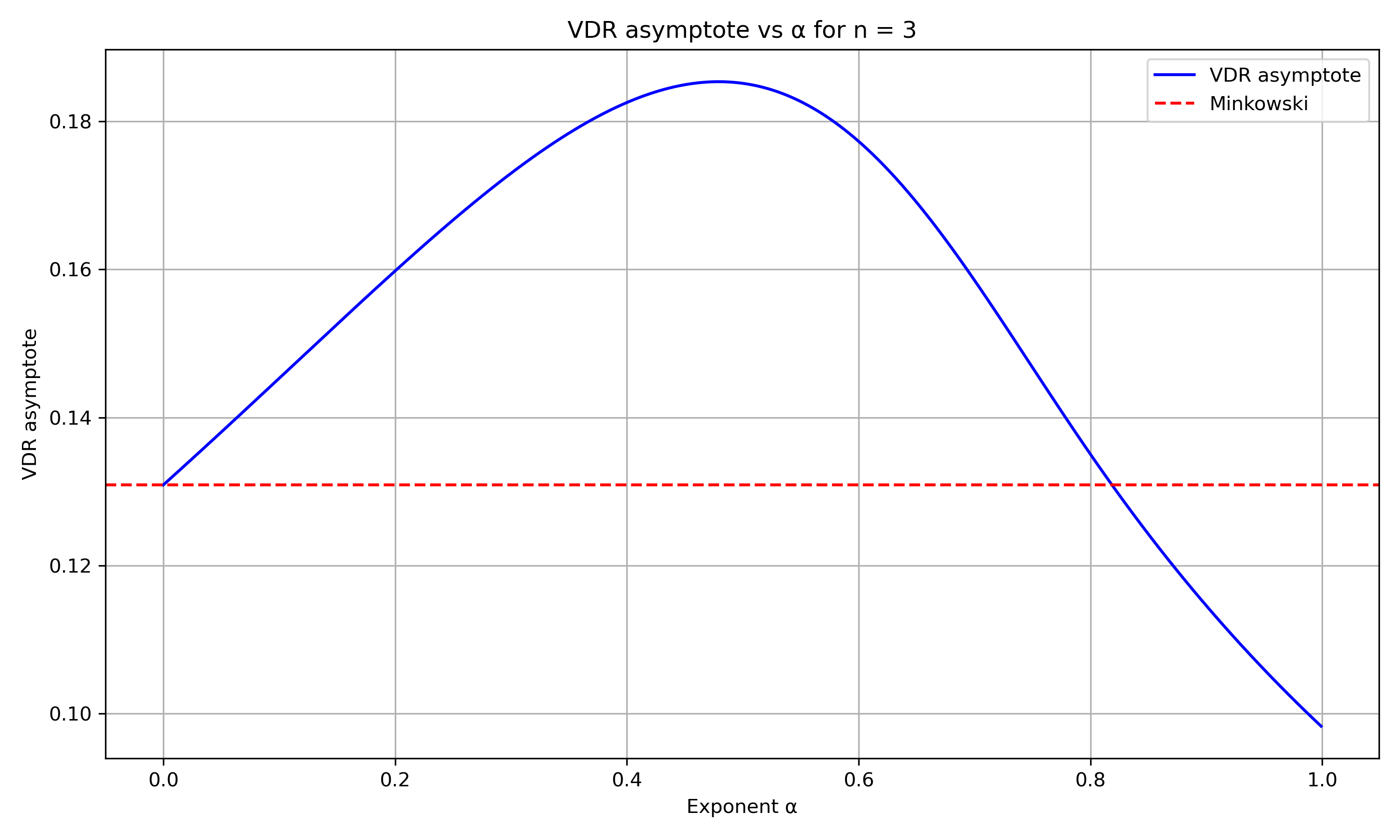}
		\includegraphics[width=0.45\textwidth]{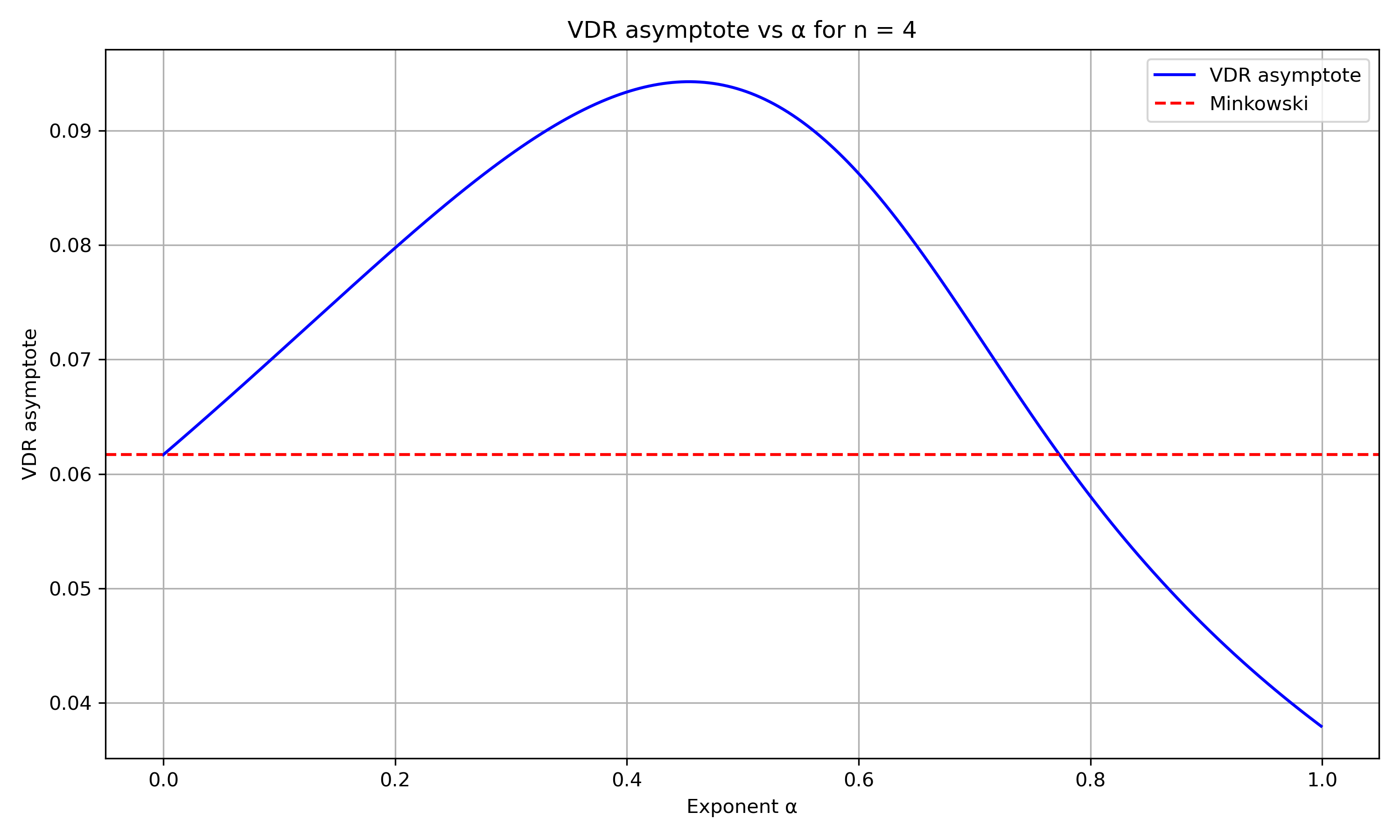}
		\caption{$R_n(\alpha)/\omega_n$ for $n=1,2,3,4$}
		\label{fig 3}
	\end{figure}
	\begin{figure}[h]
		\centering
		\includegraphics[width=0.9\textwidth]{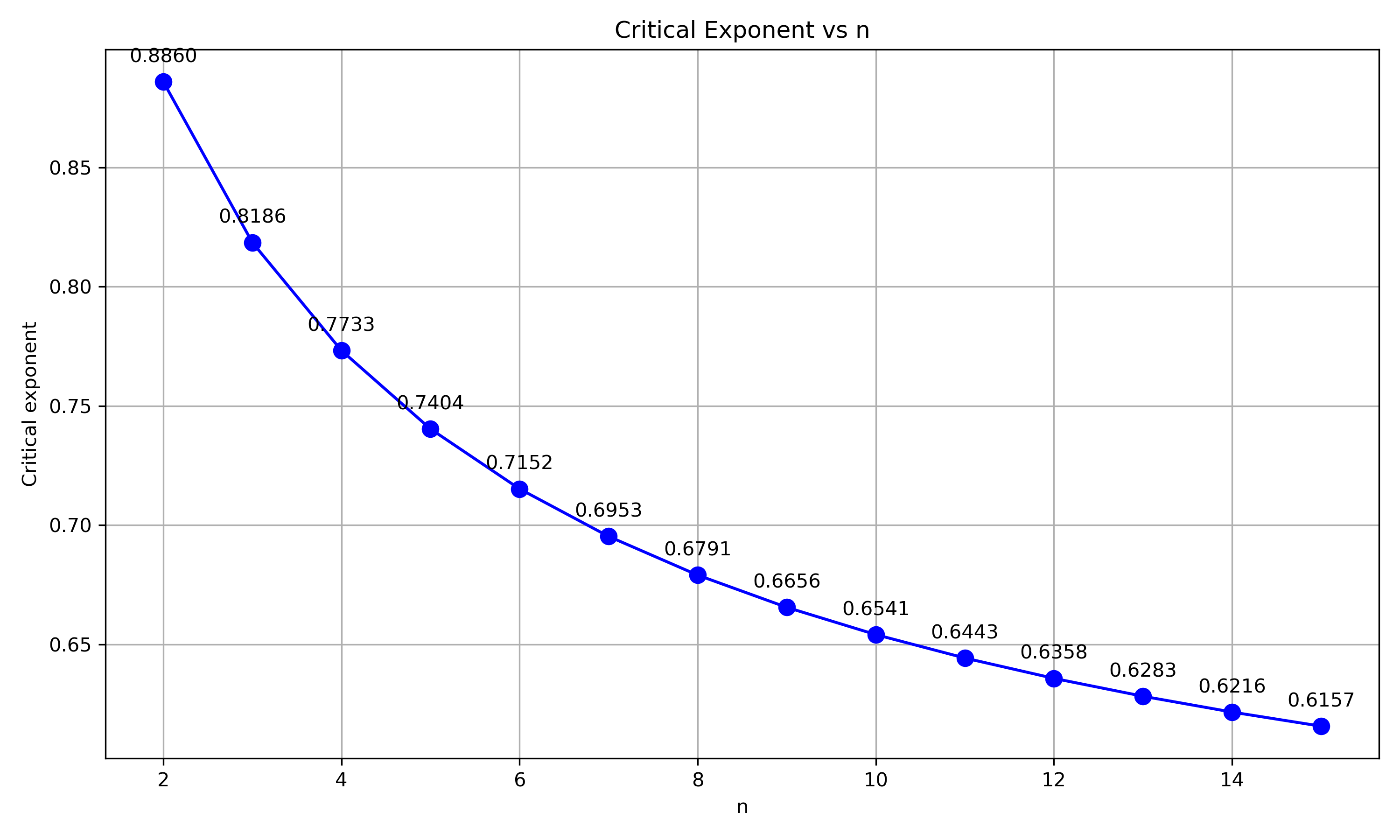}
		\caption{Critical exponents for $2\leq n \leq 15$}
		\label{fig 4}
	\end{figure}
\end{remark}

%%%%%%%%%%%%%%%%%%%%
%%%%%%%%%%%%%%%%%%%%
%%%%%%%%%%%%%%%%%%%%
\section{Spatially hyperbolic FLRW spacetimes with scale factor $a(t) \sim a_0 t^{\alpha}$}

We consider the spatially hyperbolic FLRW spacetime $(M=\mathbb{R}^{n+1}_{t>0}, g)$, with $n \geq 2$, and scale factor $a(t) \sim a_0 t^{\alpha}$
\begin{align*}
	g
	=
	- \ed t^2 + a(t)^2 ( \ed r^2 + \sinh^2 r \ringg).
\end{align*}
For explicit calculations, we consider the specific metric with scale factor $a(t) = a_0 t^{\alpha}$
\begin{align*}
	g_{a_0, \alpha, h}
	=
	- \ed t^2 + a_0^2 t^{2\alpha} ( \ed r^2 + \sinh^2 r \ringg).
\end{align*}

%%%%%%%%%%%%%%%%%%%%
%%%%%%%%%%%%%%%%%%%%
\subsection{Blow-up limit at past timelike boundary}

To study the blow-up limit at the past timelike boundary, we introduce the rescaled time $t = T \tilde{t}$.
Then the metric becomes
\begin{align*}
	g_{a_0, 1, h}
	=
	T^2
	[ - \ed \tilde{t}^2
	+ a_0^2 T^{2(\alpha -1)} \tilde{t}^{2 \alpha} ( \ed r^2 + \sinh^2 r \ringg )].
\end{align*}
Next, we define the rescaled radius $\tilde{r}$
\begin{align*}
	 &
	r
	=
	\lambda \tilde{r},
	\quad
	\lambda
	=
	a_0^{-1} T^{ 1-\alpha }.
	\\
	 &
	T^{-2} g_{a_0, \alpha, h}
	=
	- \ed \tilde{t}^2
	+ \tilde{t}^{2 \alpha}
	\big[ \ed \tilde{r}^2
		+ \big( \frac{\sinh \lambda \tilde{r}}{\lambda} )^2 \ringg
		\big].
\end{align*}
We now analyze the limit of the rescaled metric $T^{-2} g_{a_0, \alpha, h}$ as $T \rightarrow 0^+$.

%%%%%%%%%%%%%%%%%%%%
\paragraph*{Case $\alpha >1 $:}
As $T \rightarrow 0^+$,
\begin{align*}
	\lambda
	=
	a_0^{-1} T^{ 1-\alpha }
	\rightarrow
	+ \infty,
	\quad
	\frac{\sinh \lambda \tilde{r}}{\lambda}
	\sim
	\frac{1}{2\lambda} e^{\lambda \tilde{r}}
	\rightarrow
	+\infty,
	\quad
	\tilde{r} \neq 0.
\end{align*}
Hence, for $\alpha>1$, the blow-up limit of $g_{a_0, \alpha, h}$ does not exist. Formally, the limit is
\begin{align}
	- \ed \tilde{t}^2
	+ \tilde{t}^{2 \alpha}
	( \ed \tilde{r}^2
	+ \infty \ringg ).
	\label{eqn 3.1R}
\end{align}

%%%%%%%%%%%%%%%%%%%%
\paragraph*{Case $\alpha = 1$, $a_0>0$:}
The metric $g_{a_0, \alpha=1, h}$ is scale invariant, as seen from
\begin{align*}
	g_{a_0, 1, h}
	=
	T^2
	[ - \ed \tilde{t}^2
	+ a_0^2 \tilde{t}^2 ( \ed r^2 + \sinh^2 r \ringg )].
\end{align*}
Consequently, the blow-up limit of $g_{a_0,1,h}$ at the past timelike boundary point $\mathcal{O}$ is itself.

%%%%%%%%%%%%%%%%%%%%
\paragraph*{Case $\alpha \in (0,1)$:}
As $T \rightarrow 0^+$,
\begin{align*}
	\lambda
	=
	a_0^{-1} T^{ 1-\alpha }
	\rightarrow
	0^+,
	\quad
	\frac{\sinh \lambda \tilde{r}}{\lambda}
	\rightarrow
	\tilde{r}.
\end{align*}
Thus in the rescaled coordinate system $(\tilde{t}, \tilde{r}, \vartheta)$
\begin{align}
	T^{-2} g_{a_0, \alpha, h}
	\rightarrow
	g_{\alpha, f}
	= - \ed \tilde{t}^2
	+ \tilde{t}^{2 \alpha}
	( \ed \tilde{r}^2
	+ \tilde{r}^2 \ringg ).
	\label{eqn 3.2R}
\end{align}
Hence, for $\alpha \in (0,1)$, the blow-up limit of $g_{a_0, \alpha, h}$ is the corresponding spatially flat FLRW metric $g_{\alpha, f}$.

%%%%%%%%%%%%%%%%
%%%%%%%%%%%%%%%%
\subsection{Case $\alpha>1$}

%%%%%%%%%%%%%%%%
\subsubsection{Conformal compactification and past boundary}

Following the construction of the conformal compactification, define
\begin{align*}
	 &
	\tilde{t}
	=
	\int_1^t \frac{1}{a(t')} \ed t'
	\overset{t\rightarrow 0}{\longrightarrow}
	+\infty,
	\quad
	\ed \tilde{t}
	=
	\frac{\ed t}{a(t)}.
	\\
	 &
	u = \frac{\tilde{t}-r}{2},
	\quad
	v= \frac{\tilde{t}+r}{2}.
	\\
	 &
	\sinh u = \frac{\tilde{u}}{\sqrt{1-\tilde{u}^2}},
	\quad
	\sinh v = \frac{\tilde{v}}{\sqrt{1-\tilde{v}^2}}.
\end{align*}
Then
\begin{align*}
	g
	=
	\frac{a^2(t)}{(1-\tilde{u}^2)(1-\tilde{v}^2)}
	[ - 4 \ed \tilde{u} \ed \tilde{v} + (\tilde{v} - \tilde{u})^2 \ringg ].
\end{align*}
Thus, $(M,g)$ has a past timelike boundary point $\mathcal{O}$ and a past null boundary $\mathcal{PN}$:
\begin{enumerate}[label=\alph*.]
	\item
	      $\mathcal{O}$: $\tilde{u} = \tilde{v} = -1$,

	\item
	      $\mathcal{PN}$: $\tilde{u} = -1$, $\tilde{v} \in (-1,1)$.

\end{enumerate}

%%%%%%%%%%%%%%%%%%%%
\subsubsection{VDR asymptote and inextendibility at $\mathcal{PT}$}

\begin{theorem}\label{thm 3.1}
	The VDR asymptote at the past timelike boundary point $\mathcal{O}$ of the spatially hyperbolic FLRW spacetime $(M = \mathbb{R}^{n+1}_{t>0},g)$, with $n\geq 2, a_0>0, \alpha > 1$, is $+\infty$, and thus the spacetime admits {\bf no local $\mathrm{C}^0$ locally null-non-accumulating strongly-causal} or {\bf $\mathrm{C}^{0,1}$ strongly-causal extension} at $\mathcal{O}$.
\end{theorem}

\begin{proof}
	By spatial homogeneity, it suffices to calculate the VDR asymptote along any sequence $p_k = (t_k, r=0)$, $t_k \rightarrow 0$.

	Recall the blow-up limit \eqref{eqn 3.1R}:
	\begin{align*}
		 &
		t = t_k \tilde{t},
		\quad
		r = \lambda_k \tilde{r},
		\quad
		\lambda_k = a_0^{-1} t_k^{1-\alpha}
		\rightarrow +\infty.
		\\
		 &
		t_k^{-2} g
		=
		-\ed \tilde{t}
		+ \tilde{a}^2_k
		\big[ \ed \tilde{r}^2
			+ \big( \frac{\sinh \lambda_k \tilde{r}}{\lambda_k} )^2 \ringg
			\big],
		\quad
		\tilde{a}_k(\tilde{t})
		=
		\frac{a ( t_k \tilde{t})}{a_0 t_k^{\alpha}}
		\rightarrow
		\tilde{t}^{\alpha},
		\quad
		\frac{\sinh \lambda_k \tilde{r}}{\lambda_k}
		\rightarrow
		+\infty.
	\end{align*}
	Introduce the radius functions $\tilde{r}^-_k$ and $r^-_{\alpha}$ by
	\begin{align*}
		 &
		\tilde{r}^-_k ( \tilde{t} )
		=
		\int_{\tilde{t}}^1
		\frac{1}{\tilde{a}_k (\tilde{t}')}
		\ed \tilde{t}',
		\quad
		r^-_{\alpha}( t )
		=
		\int_{t}^1
		\frac{1}{(t')^{\alpha}}
		\ed t',
	\end{align*}
	so that $\tilde{r}^-_k \rightarrow r^-_{\alpha}$ as $k\rightarrow +\infty$.
	The chronological diamond $I(o,p_k)$ is given by
	\begin{align*}
		I(o,p_k)
		=
		\{ (\tilde{t}, \tilde{r}): \tilde{r} < \tilde{r}^-_k (\tilde{t}) \}.
	\end{align*}
	Thus the volume of $I(o,p_k)$ is
	\begin{align*}
		\vert I(o,p_k) \vert_g
		=
		t_k^{n+1}
		\int_0^1 \int_0^{\tilde{r}_k(\tilde{t})}
		n\omega_n \big( \frac{\sinh \lambda_k \tilde{r}}{\lambda_k} \big)^{n-1}
		\cdot
		\tilde{a}^{n}(\tilde{t})
		\ed \tilde{r} \ed \tilde{t}.
	\end{align*}
	By $\frac{\sinh \lambda_k \tilde{r}}{\lambda_k} \rightarrow +\infty$ as $k \rightarrow +\infty$, we obtain,
	\begin{align*}
		\frac{\vert I(o,p_k) \vert_g}{[d(o,p_k)]^{n+1}}
		\rightarrow
		+\infty.
	\end{align*}
	The theorem follows.
\end{proof}

%%%%%%%%%%%%%%%%%%%%
%%%%%%%%%%%%%%%%%%%%
\subsection{Case $\alpha=1$}
\label{sec 3.2}

%%%%%%%%%%%%%%%%%%%%
\subsubsection{Self-similar coordinates, conformal compactification and past boundary}
Since $g_{a_0,1,h}$ is scale invariant, we introduce a self-similar coordinate system. For the general case $a(t) \sim a_0 t$, we define
\begin{align*}
	 &
	\int_1^t \frac{a_0}{a(t')} \ed t'
	=
	\log \tilde{t},
	\quad
	\frac{a_0}{a(t)} \ed t
	=
	\frac{\ed \tilde{t} }{\tilde{t}},
	\quad
	\tilde{r} = a_0 r.
\end{align*}
Then
\begin{align*}
	\frac{(a_0 \tilde{t})^2}{a^2 (t)} g
	=
	- \ed \tilde{t}^2
	+ \tilde{t}^2 (\ed \tilde{r}^2 + a_0^2 \sinh^2 \frac{\tilde{r}}{a_0} \ringg )
\end{align*}
Introducing the transformation
$\bar{t} = \tilde{t} \cosh \tilde{r},
	\bar{r} = \tilde{t} \sinh \tilde{r}$,
we obtain
\begin{align*}
	\frac{(a_0 \tilde{t})^2}{a^2 (t)} g
	=
	- \ed \bar{t}^2
	+ \ed \bar{r}^2
	+ \big( \frac{a_0}{\sinh \tilde{r}} \sinh \frac{\tilde{r}}{a_0} \big)^2 \bar{r}^2 \ringg.
\end{align*}
Following \cite{Chr94}, we define the self-similar coordinates $u, x$ by
\begin{align*}
	\bar{t} = u(1-x),
	\quad
	\bar{r} = u x,
	\quad
	\Rightarrow
	\quad
	\ed \bar{t} = (1-x) \ed u - u \ed x,
	\quad
	\ed \bar{r} = x \ed u + u \ed x.
\end{align*}
This yields
\begin{align}
	 &
	\frac{(a_0 \tilde{t})^2}{a^2 (t)} g
	=
	- (1- 2x) \ed u^2 + 2u \ed u \ed x
	+ f^2(a_0, x) (ux)^2 \ringg,
	\label{eqn 3.3}
	\\
	 &
	f(a_0, x)
	=
	\frac{a_0}{\sinh \tilde{r}} \sinh \frac{\tilde{r}}{a_0},
	\quad
	\tanh \tilde{r}
	= \frac{\bar{r}}{\bar{t}}
	= \frac{x}{1-x},
	\quad
	x \in (0,\frac{1}{2}).
	\nonumber
\end{align}
For the factor $f(a_0, x)$ with $x \in (0,\frac{1}{2})$, we have
\begin{align*}
	 &
	a_0 = 1,
	\quad
	f(a_0=1, x) = 1,
	\\
	 &
	a_0> 1,
	\quad
	f(a_0>1, x) < 1,
	\quad
	\lim_{x \rightarrow \frac{1}{2}} f(a_0>1 ,x) = 0,
	\\
	 &
	a_0 < 1,
	\quad
	f(a_0<1, x) > 1,
	\quad
	\lim_{x \rightarrow \frac{1}{2}} f(a_0<1, x) = + \infty.
\end{align*}
The past boundary of $(M,g)$ consists of two components: the scaling origin $\mathcal{O}$ as a past timelike boundary point, and the past null boundary $\mathcal{PN}$.
\begin{enumerate}[label=\alph*.]
	\item
	      $\mathcal{O}$: $u=0$

	\item
	      $\mathcal{PN}$: $x=\frac{1}{2}$, $u>0$.

\end{enumerate}

%%%%%%%%%%%%%%%%%%%%
\subsubsection{Volume form estimate in self-similar coordinate system}

We compute the VDR asymptote explicitly in $(M,g_{a_0, 1, h})$. By scale invariance and spatial homogeneity, the VDR calculation at a single point suffices. Following \cite{Le25}, we estimate the volume form in self-similar coordinates.

Let $p = (t=1, |x| = 0)$, corresponding to $(u=1,x=0)$ in self-similar coordinates $(u, x, \vartheta)$. The chronological diamond is
\begin{align*}
	I(\mathcal{O},p)
	=
	\{ u \in (0,1), x \in (0, \frac{1}{2}) \}.
\end{align*}
The volume form of $g_{a_0,1,h}$ in these coordinates is
\begin{align*}
	\dvol_{a_0,1,h}
	=
	2 u^n x^{n-1} f^{n-1}(a_0, x) \ed u \ed x \dvol_{\ringg}.
\end{align*}
Hence,
\begin{align*}
	 &
	a_0 = 1,
	\quad
	\dvol_{1,1,h}
	=
	2 u^n x^{n-1} \ed u \ed x \dvol_{\ringg}
	=
	\dvol_{\eta},
	\\
	 &
	a_0> 1,
	\quad
	\dvol_{a_0,1,h}
	<
	\dvol_{\eta},
	\\
	 &
	a_0 < 1,
	\quad
	\dvol_{a_0,1,h}
	>
	\dvol_{\eta}.
\end{align*}
Integrating gives the volume of $I(\mathcal{O}, p)$:
\begin{align}
	\begin{aligned}
		 &
		a_0 = 1,
		\quad
		\vert I(\mathcal{O}, p) \vert_{g_{1,1,h}}
		=
		\vert I(\mathcal{O}, p) \vert_{\eta}
		=
		\frac{2\omega_n}{n+1} \cdot (\frac{1}{2})^{n+1},
		\\
		 &
		a_0> 1,
		\quad
		\vert I(\mathcal{O}, p) \vert_{g_{a_0,1,h}}
		<
		\vert I(\mathcal{O}, p) \vert_{\eta},
		\\
		 &
		a_0 < 1,
		\quad
		\vert I(\mathcal{O}, p) \vert_{g_{a_0,1,h}}
		>
		\vert I(\mathcal{O}, p) \vert_{\eta}.
	\end{aligned}
	\label{eqn 3.4}
\end{align}

For $a_0\leq 1$, the function $f(a_0,x)$ is monotonically decreasing in $a_0$ when $x>0$. Its derivative is
\begin{align*}
	\partial_{a_0} f(a_0, x)
	=
	\frac{\sinh \frac{\tilde{r}}{a_0}}{a_0 \sinh \tilde{r}}
	( a_0
	-
	\coth \frac{\tilde{r}}{a_0})
	<
	0,
	\quad
	x>0,
	\quad
	a_0 \leq 1.
\end{align*}
Thus, the volume form $\dvol_{a_0\leq 1,1,h}$ is decreasing in $a_0$ for $x>0$, and so is the volume $\vert I(\mathcal{O}, p) \vert_{g_{a_0\leq 1,1,h}}$.

%%%%%%%%%%%%%%%%%%%%
\subsubsection{VDR asymptote and inextendibility at $\mathcal{O}$}
We state the VDR asymptote and inextendibility result for the spatially hyperbolic FLRW spacetime with scale factor $a(t) \sim a_0 t$.
\begin{theorem}\label{thm 3.2}
	For spatially hyperbolic FLRW spacetime $(M = \mathbb{R}^{n+1}_{t>0}, g)$ with $n\geq 2$ and scale factor $a(t) \sim a_0 t$, depending on the value of $a_0$, three cases arise:
	\begin{enumerate}[label=\textbullet]
		\item
		      If $a_0=1$, then the VDR asymptote at $\mathcal{O}$ {\bf equals} the Minkowski value $\frac{2\omega_n}{n+1} \cdot (\frac{1}{2})^{n+1}$.

		\item
		      If $a_0>1$, then the VDR asymptote at $\mathcal{O}$ is {\bf less than} the Minkowski value $\frac{2\omega_n}{n+1} \cdot (\frac{1}{2})^{n+1}$. Moreover, $(M,g)$ admits {\bf no local $\mathrm{C}^0$ locally null-non-accumulating strongly-causal} or {\bf $\mathrm{C}^{0,1}$ strongly-causal extension} at $\mathcal{O}$.

		\item
		      If $a_0 \in (0,1)$, then the VDR asymptote at $\mathcal{O}$ is {\bf greater than} the Minkowski value $\frac{2\omega_n}{n+1} \cdot (\frac{1}{2})^{n+1}$. Furthermore, this asymptote decreases monotonically to $\frac{2\omega_n}{n+1} \cdot (\frac{1}{2})^{n+1}$ as $a_0$ increases to $1$. There exists {\bf no local $\mathrm{C}^0$ locally null-non-accumulating strongly-causal} or {\bf $\mathrm{C}^{0,1}$ strongly-causal extension} of $(M,g)$ at $\mathcal{O}$.

	\end{enumerate}
\end{theorem}

\begin{proof}
	The VDR asymptote at $\mathcal{O}$ follows from the explicit calculation \eqref{eqn 3.4} for $a(t) = a_0 t$. By spatial homogeneity, it suffices to estimate the VDR asymptote along the sequence $p_k = (t_k, |x|=0)$ as $t_k \rightarrow 0$, using the self-similar coordinate system.

	We modify the construction of the self-similar coordinates
	\begin{align*}
		 &
		\int_t^{t_k} \frac{a_0}{a(t')} \ed t'
		=
		\log t_k - \log \tilde{t},
		\quad
		\tilde{r} = a_0 r,
		\\
		 &
		\bar{t} = \tilde{t} \cosh \tilde{r},
		\quad
		\bar{r} = \tilde{t} \sinh \tilde{r},
		\\
		 &
		\bar{t} = u(1-x),
		\quad
		\bar{r} = u x.
	\end{align*}
	By \eqref{eqn 3.3},
	\begin{align*}
		\frac{(a_0 \tilde{t})^2}{a^2 (t)} g
		=
		- (1- 2x) \ed u^2 + 2u \ed u \ed x
		+ f^2(a_0, x) (ux)^2 \ringg.
	\end{align*}
	The self-similar coordinates of $p_k$ are $(u=t_k, x=0)$. The chronological diamond $I(\mathcal{O}, p_k)$ is
	\begin{align*}
		I(\mathcal{O}, p_k)
		=
		\{ u \in (0, t_k), x\in [0, \frac{1}{2} \}.
	\end{align*}
	Inside $I(\mathcal{O}, p_k)$, the volume form $\dvol_g$ is
	\begin{align*}
		\dvol_g
		=
		( 1 + c(t_k, t) ) \dvol_{a_0,1,h}.
	\end{align*}
	Letting $p=(t=1,|x|=0)$, the volume of $I(\mathcal{O},p_k)$ is
	\begin{align*}
		\vert I(\mathcal{O},p_k) \vert
		=
		(1+c(t_k)) t_k^{n+1} \vert I(\mathcal{O},p) \vert_{a_0,1,h}.
	\end{align*}
	Hence the VDR asymptote at $\mathcal{O}$ is $\vert I(\mathcal{O},p) \vert_{a_0,1,h}$. The result follows by \eqref{eqn 3.4} and Theorems \ref{thm 1.6}, \ref{thm 1.7}.
\end{proof}

%%%%%%%%%%%%%%%%%%%%
%%%%%%%%%%%%%%%%%%%%
\subsection{Case $\alpha \in (0,1)$}

%%%%%%%%%%%%%%%%%%%%
\subsubsection{Conformal compactification and past boundary}\label{sec 3.4.1}
We define the new time coordinate $\tilde{t} = \int_0^t \frac{1}{a(t')} \ed t'$, so that $\ed \tilde{t} = \frac{\ed t}{a(t)}$, which transforms the metric into
\begin{align*}
	g
	=
	a(t)^2 ( - \ed \tilde{t}^2 + \ed r^2 + \sinh^2 r \ringg ),
	\quad
	\tilde{t}>0.
\end{align*}
Introducing double null coordinates $u = \frac{\tilde{t}-r}{2}, v = \frac{\tilde{t}+r}{2}$, $u+v>0, v-u>0$, we obtain
\begin{align*}
	g
	=
	a^2(t) [ - 4 \ed u \ed v + \sinh^2 (v-u) \ringg ].
\end{align*}
To construct the conformal compactification, define $\tilde{u}, \tilde{v}$ via
\begin{align*}
	 &
	\sinh u = \frac{\tilde{u}}{\sqrt{1-\tilde{u}^2}},
	\quad
	\sinh v = \frac{\tilde{v}}{\sqrt{1-\tilde{v}^2}},
	\quad
	\sinh(v-u)
	=
	\frac{\tilde{v} - \tilde{u}}{\sqrt{1-\tilde{u}^2} \cdot\sqrt{1-\tilde{v}^2}},
	\\
	 &
	\ed u = \frac{\ed \tilde{u}}{1-\tilde{u}^2},
	\quad
	\ed v = \frac{\ed \tilde{v}}{1-\tilde{v}^2},
\end{align*}
which yields the metric
\begin{align*}
	g
	=
	\frac{a^2(t)}{(1-\tilde{u}^2)(1-\tilde{v}^2)}
	[ - 4 \ed \tilde{u} \ed \tilde{v} + (\tilde{v} - \tilde{u})^2 \ringg ].
\end{align*}
Thus, the spacetime $(M,g)$ possesses a past timelike boundary $\mathcal{PT}$ and no past null boundary, given by
\begin{align*}
	\mathcal{PT}
	=
	\{ \tilde{u} + \tilde{v} = 0 \}
	=
	\{ \tilde{t} = 0 \}
	=
	\{ t = 0 \}.
\end{align*}

%%%%%%%%%%%%%%%%%%%%
\subsubsection{VDR asymptote and inextendibility at $\mathcal{PT}$}\label{sec 3.4.2}

We compute the VDR asymptote along the $\Sigma_t$-orthogonal geodesic using the blow-up limit \eqref{eqn 3.2R}.

\begin{definition}
	For the spatially hyperbolic FLRW spacetime $(M=\mathbb{R}^{n+1}_{t>0}, g)$, $n\geq 2$ with scale factor $a(t) \sim a_0 t^{\alpha}$, $\alpha\in(0,1)$, we call an exponent $\alpha$ critical if the VDR asymptote along the $\Sigma_t$-orthogonal geodesic $\gamma(t) = (t,x)$ at the past timelike boundary point $(t=0,x)$ {\bf equals} the Minkowski value $ \frac{2\omega_n}{n+1} \cdot (\frac{1}{2})^{n+1}$.
\end{definition}

\begin{theorem}\label{thm 3.4}
	For the spatially hyperbolic FLRW spacetime $(M = \mathbb{R}^{n+1}_{t>0},g)$, with $n\geq 2, a_0>0, \alpha \in (0,1)$, an exponent is critical if and only if it is critical for the spatially flat FLRW spacetime with scale factor $t^{\alpha}$, since the VDR asymptote along the $\Sigma_t$-orthogonal geodesic $\gamma(t) = (t,x)$ at the past timelike boundary point $(t=0,x)$ equals the corresponding asymptote in the spatially flat case; consequently, the spacetime admits {\bf no local $\mathrm{C}^{0,1}$ strongly-causal extension} at any past timelike boundary point for a non-critical exponent $\alpha$.
\end{theorem}

\begin{proof}
	By spatial homogeneity, it suffices to calculate the VDR asymptote along $\gamma(t) = (t, r=0)$ at the past timelike point $o=(t=0,r=0)$.

	Recall the blow-up limit \eqref{eqn 3.2R}:
	\begin{align*}
		 &
		t = T \tilde{t},
		\quad
		T\rightarrow 0,
		\quad
		r = \lambda_T \tilde{r},
		\quad
		\lambda_T = a_0^{-1} T^{1-\alpha}
		\rightarrow 0.
		\\
		 &
		T^{-2} g
		=
		-\ed \tilde{t}
		+ \tilde{a}^2_T
		\big[ \ed \tilde{r}^2
			+ \big( \frac{\sinh \lambda_T \tilde{r}}{\lambda_T} )^2 \ringg
			\big],
		\quad
		\tilde{a}_T(\tilde{t})
		=
		\frac{a ( T \tilde{t})}{a_0 T^{\alpha}}.
	\end{align*}
	Define $p_T = (t=T, r=0) = (\tilde{t}=1, \tilde{r}=0)$ and introduce the radius functions $\tilde{r}^{\pm}_T$ by
	\begin{align*}
		\tilde{r}^+_T( \tilde{t} )
		=
		\int_0^{\tilde{t}}
		\frac{1}{\tilde{a}_T(\tilde{t}')}
		\ed \tilde{t}',
		\quad
		\tilde{r}^-_T( \tilde{t} )
		=
		\int_{\tilde{t}}^1
		\frac{1}{\tilde{a}_T(\tilde{t}')}
		\ed \tilde{t}'.
		\quad
		\tilde{r}_T (\tilde{t}) = \min \{ \tilde{r}^{\pm}_T(\tilde{t}) \},
		\quad
		\tilde{t} \in [0,1].
	\end{align*}
	The chronological diamond $I(o,p_T)$ for the metric $g$ is
	\begin{align*}
		I(o,p_T)
		=
		\{ (\tilde{t}, \tilde{r}): \tilde{r} < \tilde{r}_T(\tilde{t}) \}.
	\end{align*}
	Thus the volume of $I(o,p_T)$ is given by
	\begin{align*}
		\vert I(o,p_T) \vert_g
		=
		T^{n+1}
		\int_0^1 \int_0^{\tilde{r}_T(\tilde{t})}
		n\omega_n \big( \frac{\sinh \lambda_T \tilde{r}}{\lambda_T} \big)^{n-1}
		\cdot
		\tilde{a}^{n}(\tilde{t})
		\ed \tilde{r} \ed \tilde{t}
	\end{align*}

	Recall the radius functions $r_{\alpha}^{\pm}$ for the chronological diamond $I_{\alpha,f}(o,p), p=(t=1,r=0)$ in the spatially flat case in section \ref{sec 2.2.2}
	\begin{align*}
		 &
		r_{\alpha}^+( t )
		=
		\int_0^{t}
		\frac{1}{(t')^{2\alpha}}
		\ed t',
		\quad
		r_{\alpha}^-( t )
		=
		\int_{t}^1
		\frac{1}{(t')^{2\alpha}}
		\ed t'.
		\quad
		r_{\alpha} (t) = \min \{ r_{\alpha}^{\pm}(t) \},
		\\
		 &
		I_{\alpha,f}(o,p)
		=
		\{ (t, r): r < r_{\alpha} (t) \},
	\end{align*}
	and the volume $\vert I_{\alpha,f}(o,p) \vert_{g_{\alpha,f}}$ is
	\begin{align*}
		\vert I_{\alpha,f}(o,p) \vert_{g_{\alpha,f}}
		=
		\int_0^1 \int_0^{r_{\alpha,f} (t)}
		n\omega_n r^{n-1}
		\cdot
		(t^{\alpha} )^n
		\ed r \ed t.
	\end{align*}
	Since $\tilde{r}_T \rightarrow r_{\alpha}$, $ \frac{\sinh \lambda_T \tilde{r}}{\lambda_T} \rightarrow \tilde{r} $ as $T\rightarrow 0$, we obtain
	\begin{align*}
		\frac{\vert I(o,p_T) \vert_g}{d(o,p_T)^{n+1}}
		=
		\frac{\vert I(o,p_T) \vert_g}{T^{n+1}}
		\rightarrow
		\vert I_{\alpha,f}(o,p) \vert_{g_{\alpha,f}}.
	\end{align*}
	The theorem follows.
\end{proof}

%%%%%%%%%%%%%%%%%%%%
%%%%%%%%%%%%%%%%%%%%
%%%%%%%%%%%%%%%%%%%%
\section{Spatially spherical FLRW spacetimes with scale factor $a(t) \sim a_0 t^{\alpha}$}

We consider the spatially spherical FLRW spacetime $(M=\mathbb{R}_+ \times\mathbb{S}^n, g)$, with $n \geq 1$, and scale factor $a(t) \sim a_0 t^{\alpha}$
\begin{align*}
	g
	=
	- \ed t^2 + a(t)^2 ( \ed r^2 + \sin^2 r \ringg),
	\quad
	r\in [0,\pi).
\end{align*}
For explicit calculations, we consider the specific metric with scale factor $a(t) = a_0 t^{\alpha}$
\begin{align*}
	g_{a_0, \alpha, s}
	=
	- \ed t^2 + a_0^2 t^{2\alpha} ( \ed r^2 + \sin^2 r \ringg).
\end{align*}

%%%%%%%%%%%%%%%%%%%%
%%%%%%%%%%%%%%%%%%%%
\subsection{Blow-up limit at past timelike boundary}

Following the blow-up limit construction for the spatially hyperbolic case, we introduce the rescaled time $\tilde{t}$ and rescaled radius $\tilde{r}$:
\begin{align*}
	 &
	t = T \tilde{t},
	\quad
	r = \lambda \tilde{r},
	\quad
	\lambda = a_0^{-1} T^{ 1-\alpha },
	\\
	 &
	T^{-2} g_{a_0, 1, s}
	=
	- \ed \tilde{t}^2
	+ a_0^2 T^{2(\alpha -1)} \tilde{t}^{2 \alpha} ( \ed r^2 + \sin^2 r \ringg )
	\\
	 & \phantom{T^{-2} g_{a_0, 1, s}}
	=
	- \ed \tilde{t}^2
	+ \tilde{t}^{2 \alpha}
	\big[ \ed \tilde{r}^2
		+ \big( \frac{\sin \lambda \tilde{r}}{\lambda} )^2 \ringg
		\big].
\end{align*}
We now analyze the limit of the rescaled metric $T^{-2} g_{a_0, \alpha, s}$ as $T \rightarrow 0^+$.

%%%%%%%%%%%%%%%%%%%%
\paragraph*{Case $\alpha >1 $:}
As $T \rightarrow 0^+$,
\begin{align*}
	\lambda
	=
	a_0^{-1} T^{ 1-\alpha }
	\rightarrow
	+ \infty,
	\quad
	\tilde{r}
	\in
	[0, \lambda^{-1} \pi),
	\quad
	\frac{\sin \lambda \tilde{r}}{\lambda}
	\rightarrow
	0.
\end{align*}
Hence, for $\alpha>1$, the blow-up limit of $g_{a_0, \alpha, s}$ does not exist.

%%%%%%%%%%%%%%%%%%%%
\paragraph*{Case $\alpha = 1$, $a_0>0$:}
The metric $g_{a_0, 1, s}$ is scale invariant, as seen from
\begin{align*}
	g_{a_0, 1, s}
	=
	T^2
	[ - \ed \tilde{t}^2
	+ a_0^2 \tilde{t}^2 ( \ed r^2 + \sin^2 r \ringg )].
\end{align*}
Consequently, the blow-up limit of $g_{a_0,1,s}$ at the past timelike boundary point $\mathcal{O}$ is itself.

%%%%%%%%%%%%%%%%%%%%
\paragraph*{Case $\alpha \in (0,1)$:}
As $T \rightarrow 0^+$,
\begin{align*}
	\lambda
	=
	a_0^{-1} T^{ 1-\alpha }
	\rightarrow
	0^+,
	\quad
	\tilde{r}
	\in
	[0, \lambda^{-1} \pi),
	\quad
	\frac{\sin \lambda \tilde{r}}{\lambda}
	\rightarrow
	\tilde{r}.
\end{align*}
Thus in the rescaled coordinate system $(\tilde{t}, \tilde{r}, \vartheta)$
\begin{align}
	T^{-2} g_{a_0, \alpha, s}
	\rightarrow
	g_{\alpha, f}
	= - \ed \tilde{t}^2
	+ \tilde{t}^{2 \alpha}
	( \ed \tilde{r}^2
	+ \tilde{r}^2 \ringg ).
	\label{eqn 4.1}
\end{align}
Hence, for $\alpha \in (0,1)$, the blow-up limit of $g_{a_0, \alpha, s}$ is the corresponding spatially flat FLRW metric $g_{\alpha, f}$.

%%%%%%%%%%%%%%%%
%%%%%%%%%%%%%%%%
\subsection{Case $\alpha>1$}

The past boundary of $(M,g)$ includes a past timelike boundary point $\mathcal{O}=\{t=0\}$. We have the following result on the VDR asymptote and inextendibility at $\mathcal{O}$.

\begin{theorem}
	The VDR asymptote at the past timelike boundary point $\mathcal{O}$ of the spatially spherical FLRW spacetime $(M = \mathbb{R}_+ \times\mathbb{S}^n,g)$, with $n\geq 1, a_0>0, \alpha > 1$, is $0$. The spacetime admits {\bf no local $\mathrm{C}^0$ strongly-causal} at $\mathcal{O}$.
\end{theorem}

\begin{proof}
	By spatial homogeneity, it suffices to calculate the VDR asymptote along any sequence $p_k = (t_k, r=0)$, $t_k \rightarrow 0$.

	Recall the blow-up limit \eqref{eqn 3.1R}:
	\begin{align*}
		 &
		t = t_k \tilde{t},
		\quad
		r = \lambda_k \tilde{r} \in [0,\pi),
		\quad
		\lambda_k = a_0^{-1} t_k^{1-\alpha}
		\rightarrow +\infty.
		\\
		 &
		t_k^{-2} g
		=
		-\ed \tilde{t}
		+ \tilde{a}^2_k
		\big[ \ed \tilde{r}^2
			+ \big( \frac{\sin \lambda_k \tilde{r}}{\lambda_k} )^2 \ringg
			\big],
		\quad
		\tilde{a}_k(\tilde{t})
		=
		\frac{a ( t_k \tilde{t})}{a_0 t_k^{\alpha}}
		\rightarrow
		\tilde{t}^{\alpha},
		\quad
		\frac{\sin \lambda_k \tilde{r}}{\lambda_k}
		\rightarrow
		0.
	\end{align*}
	Following the strategy proving Theorem \ref{thm 3.1}, introduce the radius functions $\tilde{r}^-_k$ by
	\begin{align*}
		 &
		\tilde{r}^-_k ( \tilde{t} )
		=
		\max\{
		\int_{\tilde{t}}^1 \frac{1}{\tilde{a}_k (\tilde{t}')} \ed \tilde{t}' ,
		\frac{\pi}{\lambda_k}
		\},
		\quad
		\tilde{r}^-_k ( \tilde{t} ) \cdot t_k^{1-\alpha}
		\rightarrow
		a_0 \pi,
		\quad
		\tilde{r}^-_k ( \tilde{t} )
		\rightarrow
		0,
		\quad
		k \rightarrow +\infty,
		\\
		 &
		I(o,p_k)
		=
		\{ (\tilde{t}, \tilde{r}): \tilde{r} < \tilde{r}^-_k (\tilde{t}) \}.
	\end{align*}
	Thus the volume of $I(o,p_k)$ is
	\begin{align*}
		\vert I(o,p_k) \vert_g
		=
		t_k^{n+1}
		\int_0^1 \int_0^{\tilde{r}^-_k(\tilde{t})}
		n\omega_n \big( \frac{\sin \lambda_k \tilde{r}}{\lambda_k} \big)^{n-1}
		\cdot
		\tilde{a}^{n}(\tilde{t})
		\ed \tilde{r} \ed \tilde{t},
		\quad
		\frac{\vert I(o,p_k) \vert_g}{[d(o,p_k)]^{n+1}}
		\rightarrow
		0.
	\end{align*}
	Therefore, by Theorem \ref{thm 1.7}, the spacetime admits no local $\mathrm{C}^0$ locally null-non-accumulating strongly-causal extension at $\mathcal{O}$. Note that the past horismos for points with sufficiently small $t$ is compact. Hence, by \cite{Le25} Proposition 4.21, any local extension at $\mathcal{O}$ is locally null-non-accumulating. Consequently, the spacetime admits no local $\mathrm{C}^0$ strongly-causal extension at $\mathcal{O}$.
\end{proof}

%%%%%%%%%%%%%%%%%%%%
%%%%%%%%%%%%%%%%%%%%
%%%%%%%%%%%%%%%%%%%%
\subsection{Case $\alpha=1$}

We introduce a self-similar coordinate system for $g_{a_0,1,s}$. For the general case $a(t) \sim a_0 t$, define
\begin{align*}
	 &
	\int_1^t \frac{a_0}{a(t')} \ed t'
	=
	\log \tilde{t},
	\quad
	\frac{a_0}{a(t)} \ed t
	=
	\frac{\ed \tilde{t} }{\tilde{t}},
	\quad
	\tilde{r} = a_0 r \in [0, a_0 \pi).
	\\
	 &
	\frac{(a_0 \tilde{t})^2}{a^2 (t)} g
	=
	- \ed \tilde{t}^2
	+ \tilde{t}^2 (\ed \tilde{r}^2 + a_0^2 \sin^2 \frac{\tilde{r}}{a_0} \ringg )
\end{align*}
Introducing the transformation
$\bar{t} = \tilde{t} \cosh \tilde{r}, \bar{r} = \tilde{t} \sinh \tilde{r}$,
and the self-similar coordinates $u,x$ by $\bar{t} = u(1-x), \bar{r} = u x$ following \cite{Chr94},
we obtain
\begin{align*}
	 &
	\frac{(a_0 \tilde{t})^2}{a^2 (t)} g
	=
	- \ed \bar{t}^2
	+ \ed \bar{r}^2
	+ \big( \frac{a_0}{\sinh \tilde{r}} \sin \frac{\tilde{r}}{a_0} \big)^2 \bar{r}^2 \ringg,
	\\
	 &
	\frac{(a_0 \tilde{t})^2}{a^2 (t)} g
	=
	- (1- 2x) \ed u^2 + 2u \ed u \ed x
	+ h^2(a_0, x) (ux)^2 \ringg,
	\\
	 &
	h(a_0, x)
	=
	\frac{a_0}{\sinh \tilde{r}} \sin \frac{\tilde{r}}{a_0},
	\quad
	\tanh \tilde{r}
	= \frac{\bar{r}}{\bar{t}}
	= \frac{x}{1-x}
	\in (0, \tanh (a_0 \pi)).
\end{align*}
For $x>0$, the factor $h(a_0, x)$ satisfies
\begin{align*}
	h(a_0, x) < \frac{\tilde{r}}{\sinh \tilde{r}} <1.
\end{align*}
The past boundary of $(M,g)$ includes the scaling origin $\mathcal{O}$ as a past timelike boundary point: $\mathcal{O} = \{u=0\}$.

Following the approach in Section \ref{sec 3.2}, we establish the following result concerning the VDR asymptote and inextendibility at $\mathcal{O}$.
\begin{theorem}
	The spatially spherical FLRW spacetime $(M = \mathbb{R}_+\times \mathbb{S}^n,g)$, with $n\geq 1, a_0>0, \alpha =1$, admits {\bf no local $\mathrm{C}^0$ strongly-causal extension} at the past timelike boundary point $\mathcal{O}$. The VDR asymptote at $\mathcal{O}$ is {\bf less than} the Minkowski value $\frac{2\omega_n}{n+1} \cdot (\frac{1}{2})^{n+1}$.
\end{theorem}
\begin{proof}
	The estimate for the VDR asymptote at $\mathcal{O}$ follows the same method as in the proof of Theorem \ref{thm 3.2}. Since the past horismos for points with sufficiently small $t$ is compact, by Theorem \ref{thm 1.7} and \cite{Le25} Proposition 4.21, the theorem follows.
\end{proof}

%%%%%%%%%%%%%%%%%%%%
%%%%%%%%%%%%%%%%%%%%
\subsection{Case $\alpha\in (0,1)$}

Following the construction of the conformal compactification in Section \ref{sec 3.4.1}, we define
\begin{align*}
	 &
	\tilde{t} = \int_0^t \frac{1}{a(t')} \ed t',
	\quad
	\ed \tilde{t} = \frac{\ed t}{a(t)},
	\\
	 &
	\sinh \frac{\tilde{t}-r}{2} = \frac{u}{\sqrt{1-u^2}},
	\quad
	\sinh \frac{\tilde{t}+r}{2} = \frac{v}{\sqrt{1-v^2}},
	\quad
	v\in (0,1),
	\quad
	u\in (-v,v),
	\\
	 &
	\sinh \tilde{t}
	=
	\frac{v + u}{\sqrt{1-u^2} \cdot\sqrt{1-v^2}},
	\quad
	\sinh r
	=
	\frac{v - u}{\sqrt{1-u^2} \cdot\sqrt{1-v^2}},
	\\
	 &
	g
	=
	\frac{a^2(t)}{(1-u^2)(1-u^2)}
	( - 4 \ed u \ed v + \sin^2 r \cdot \ringg ).
\end{align*}
Thus $(M,g)$ possesses a past timelike boundary $\mathcal{PT}$ and no past null boundary
\begin{align*}
	\mathcal{PT}
	=
	\{ u + v = 0 \}
	=
	\{ \tilde{t} = 0 \}
	=
	\{ t = 0 \}.
\end{align*}

Following Section \ref{sec 3.4.2}, we have the following definition of critical components, VDR asymptote and inextendibility at $\mathcal{PT}$.
\begin{definition}
	For the spatially spherical FLRW spacetime $(M=\mathbb{R}_+\times \mathbb{S}^n, g)$, $n\geq 2$ with scale factor $a(t) \sim a_0 t^{\alpha}$, $\alpha\in(0,1)$, we call an exponent $\alpha$ critical if the VDR asymptote along the $\Sigma_t$-orthogonal geodesic $\gamma(t) = (t,x)$ at the past timelike boundary point $(t=0,x)$ {\bf equals} the Minkowski value $ \frac{2\omega_n}{n+1} \cdot (\frac{1}{2})^{n+1}$.
\end{definition}

\begin{theorem}
	For the spatially spherical FLRW spacetime $(M = \mathbb{R}_+\times \mathbb{S}^n,g)$, with $n\geq 2, a_0>0, \alpha \in (0,1)$, an exponent is critical if and only if it is critical for the spatially flat FLRW spacetime with scale factor $t^{\alpha}$, since the VDR asymptote along the $\Sigma_t$-orthogonal geodesic $\gamma(t) = (t,x)$ at the past timelike boundary point $(t=0,x)$ equals the corresponding asymptote in the spatially flat case; consequently, the spacetime admits {\bf no local $\mathrm{C}^{0,1}$ strongly-causal extension} at any past timelike boundary point for a non-critical exponent $\alpha$.
\end{theorem}
\begin{proof}
	The theorem follows the same argument as in Theorem \ref{thm 3.4} the case of spatially hyperbolic FLRW spacetime.
\end{proof}

For the case $n=1$, we have the following theorem.
\begin{theorem}
	The spacetime $(M = \mathbb{R}_+ \times \mathbb{S}^1,g)$ with $a_0>0$, $\alpha\in (0,1)$
	admits {\bf no local $\mathrm{C}^{0,1}$ strongly-causal} at the past timelike boundary point $\mathcal{O}$.	The VDR asymptote along a $\Sigma_t$-orthogonal geodesic $\gamma(t) = (t,x)$ approaching a past timelike boundary point is {\bf greater than} the Minkowski value $\frac{\omega_1}{4}$.
\end{theorem}
\begin{proof}
	A direct calculation analogous to that in Section \ref{sec 2.2.2} shows that the VDR asymptote along a $\Sigma_t$-orthogonal geodesic $\gamma(t) = (t,x)$ equals the one in the spacetime $(\mathbb{R}^2_{t>0},g)$ with scale factor $t^{\alpha}$, from which the theorem follows.
\end{proof}

%%%%%%%%%%%%%%%%%%%%
%%%%%%%%%%%%%%%%%%%%
%%%%%%%%%%%%%%%%%%%%
\section*{Acknowledgements}
\addcontentsline{toc}{section}{Acknowledgements}
The author is supported by the National Key R\&D Program of China 2024YFA1015000 and the National Natural Science Foundation of China 12201338.

%%%%%%%%%%%%%%%%%%%%
%%%%% APPENDIX %%%%%%%%%
%%%%%%%%%%%%%%%%%%%%

%\begin{appendices}

%%%%%%%%%%%%%%%%%%%%
%\section{}

%\end{appendices}

\Address

\end{document}